\documentclass[aps,prX,twocolumn,superscriptaddress,showpacs]{revtex4-1}
\usepackage{graphicx}
\usepackage{amsmath}
\usepackage{mathrsfs}
\usepackage{color}
\begin{document}

\title{Tuning order-by-disorder multiferroicity in CuO by doping}


\author{J. Hellsvik}
\affiliation{Dipartimento di Fisica, Universit\`a di Roma ``La Sapienza'', P. Aldo Moro 2, I-00185 Roma, Italy}
\affiliation{Istituto dei Sistemi Complessi, Consiglio Nazionale delle Ricerche, Italy}
\affiliation{Consiglio Nazionale delle Ricerche (CNR-SPIN), Via Vetoio 10, I-67100 L'Aquila, Italy}

\author{M. Balestieri}
\affiliation{Dipartimento di Fisica, Universit\`a di Roma ``La Sapienza'', P. Aldo Moro 2, I-00185 Roma, Italy}
\affiliation{Istituto dei Sistemi Complessi, Consiglio Nazionale delle Ricerche, Italy}

\author{T. Usui}
\affiliation{Division of Materials Physics, Graduate School of Engineering Science, Osaka University, Toyonaka, Osaka 560-8531, Japan}

\author{A. Stroppa}
\affiliation{Consiglio Nazionale delle Ricerche (CNR-SPIN), Via Vetoio 10, I-67100 L'Aquila, Italy}

\author{A. Bergman}
\affiliation{Department of Physics and Astronomy, Uppsala University, Box 516, SE-751~20 Uppsala, Sweden}

\author{L. Bergqvist}
\affiliation{Department of Materials and Nano Physics, KTH Royal Institute of Technology, Electrum 229, SE-164~40 Kista, Sweden}

\author{D. Prabhakaran}
\affiliation{Department of Physics, University of Oxford, Clarendon Laboratory, Parks Road, Oxford OX1 3PU, UK}

\author{O. Eriksson}
\affiliation{Department of Physics and Astronomy, Uppsala University, Box 516, SE-751~20 Uppsala, Sweden}

\author{S. Picozzi}
\affiliation{Consiglio Nazionale delle Ricerche (CNR-SPIN), Via Vetoio 10, I-67100 L'Aquila, Italy}

\author{T. Kimura}
\affiliation{Division of Materials Physics, Graduate School of Engineering Science, Osaka University, Toyonaka, Osaka 560-8531, Japan}

\author{J. Lorenzana}
\email[]{jose.lorenzana@roma1.infn.it}
\affiliation{Dipartimento di Fisica, Universit\`a di Roma ``La Sapienza'', P. Aldo Moro 2, I-00185 Roma, Italy}
\affiliation{Istituto dei Sistemi Complessi, Consiglio Nazionale delle Ricerche, Italy}

\date{\today}

\begin{abstract}
The high Curie temperature multiferroic compound, CuO, has a
quasidegenerate magnetic ground state that makes it prone to
manipulation by the so called ``order-by-disorder'' mechanism. 
First principle computations supplemented with Monte Carlo simulations
and experiments show that isovalent doping allows to
stabilize the multiferroic phase in non-ferroelectric regions of the
pristine material phase-diagram with experiments reaching a 250\%
widening of the ferroelectric temperature window with 5\% of Zn
doping. Our results allow to validate the importance of  a
quasidegenerate ground state on promoting multiferroicity on CuO at
high temperatures and open a path to the material engineering of new
multiferroic materials.  
\end{abstract}



\maketitle

\section{Introduction}\label{sec:intro}
The prospect of magnetoelectric coupling in spintronic devices, enabling the magnetic ordering to be controlled by an electric field and the ferroelectric ordering to be controlled by a magnetic field, has been a strong driving force in the research on multiferroic materials \cite{Wang09}. This class of materials is characterized by the simultaneous presence of two or more order parameters of different nature (like electric and magnetic). The high transition temperatures and the unusual sequence of the magnetic and multiferroic phases \cite{Kimura08} have been strong motivations for experimental and theoretical investigations on CuO. Recent experiments report electric field control of the chiral magnetic domains \cite{Babkevich12}, an ultrafast phase transition when exciting the system with femtosecond laser pulses \cite{Johnson12}, and the presence of a third magnetic phase revealed in ultrasonic velocity measurements and a Landau analysis\cite{Villarreal12}. 

\begin{figure}
\includegraphics[width=0.50\textwidth]{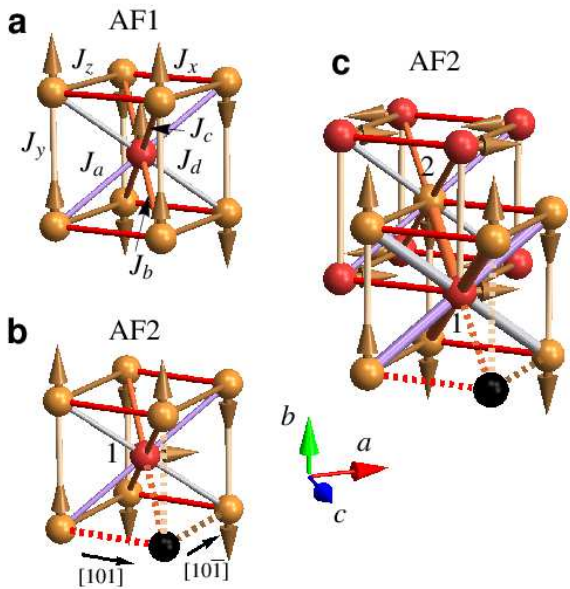}
\caption{\label{fig:atoms} Spheres represent Cu atoms in even planes (orange), odd planes (red) or impurities (black). Brown arrows represent spins. 
(a), Low-temperature AF1 state. We also show some of the exchange
interactions considered. $J_a$, $J_b$, $J_c$, $J_d$ connect sites
defined as ``first neighbors''. Bonds of the same color are equivalent
and have the same exchange. Because the Cu ion is at an inversion
center in the undistorted structure the Weiss field due to the orange
Cu's on the red Cu cancels (and viceversa). More  interactions are
shown on Fig.~\ref{fig:MB_fig_3_9}.
(b) First-rank Henley
effect. In the presence of an impurity (non-magnetic in the example)
inversion symmetry is locally broken (dashed vs. full orange bond) and
the cancellation of the Weiss field at atom 1 is no longer valid. (c)
Second-rank Henley effect. If the wave-function at atom 1 is perturbed
due to the impurity, all the intersublattice interactions indicated by
thick bonds get affected. 
}
\end{figure}

Cupric oxide crystallizes in a monoclinic structure with space group C2/c (No 15). The only point group symmetry is inversion. The lattice can be divided into two interpenetrating sublattices, one residing on planes with integer Wyckoff position along the $[010]$-direction (=$y$-axis), hereafter termed ``even planes'' and one residing on planes with half-integer Wyckoff position along $y$, termed ``odd planes'' (Fig.~\ref{fig:atoms}). The dominant magnetic interaction $J_z\sim 100$ meV produces an antiferromagnetic alignment of spins within one sublattice along $[10\bar{1}]$ while other weaker interactions tend to align same sublattice spins ferromagnetically in the $[101]$-direction and $y$-direction. 
The ground state (AF1) spin configuration is a collinear ordering with the $y$-direction as easy axis and ordering vector $\textbf{q}_{AF1}=(0.5,0, -0.5)$ (r.l.u). The multiferroic phase, AF2, occurs above the first order phase transition at $T_{\textrm{N1}}=213$~K and ranges up to a subsequent weakly first order phase transition at $T_{\textrm{N2}}=230$~K. The magnetic structure within one sublattice is similar as in the AF1 phase, but the spins on nearest neighbor planes are nearly perpendicular to each other. On top of this, the spins form a long wavelength spiral \cite{Forsyth88,Yang89,Brown91}. 

At the classical Heisenberg level, neglecting the small incommensuration, the Weiss field acting on spins of one sublattice due to the spins in the other sublattice cancel [Fig.~\ref{fig:atoms}(a)], leading to a degenerate state with an undefined angle between the spins on different sublattices \cite{Giovannetti11}. Thus state selection occurs through small perturbations \cite{Villarreal12,Giovannetti11,Jin12} like magnetic anisotropies, biquadratic terms in the Hamiltonian \cite{Kaplan09}, the order-by-disorder mechanism \cite{Henley89} and competing interactions \cite{Yablonskii90}. Monte Carlo computations have shown that presence of the spiral is essential to stabilize the nearly perpendicular configuration of spins in different sublattices \cite{Jin12} as proposed by Yablonskii long ago \cite{Yablonskii90}. On the other hand density functional theory computations have shown that the incommensuration itself is not essential to explain the magnitude of the ferroelectric moment. Indeed, the correct value of the polarization can be obtained neglecting the spiral \cite{Giovannetti11,Jin12}.

The aim of the present study is to explore how the multiferroic
properties of CuO can be tuned by isovalent doping in the Cu site
(i.e. Cu$_{1-x}$M$_x$O with M a metal ion). Henley's arguments
\cite{Henley89} suggest that impurities will stabilize the
multiferroic AF2 phase as schematically shown in
Fig.~\ref{fig:atoms}(b) (hereafter ``first-rank Henley effect''). In
the presence of an impurity (non-magnetic in the example) inversion
symmetry is locally broken (dashed vs. full orange bond) and the
cancellation of the Weiss field at atom 1 is no longer valid. Random
fields appear which are parallel to the spins on even planes and act
on the spins in odd planes (and viceversa). This leads to a
stabilization of the AF2 phase by a mechanism analogous to the one by which
the spins of an ordinary AF point in a direction approximately
perpendicular to an external field\cite{Henley89}.  

Fig.~\ref{fig:atoms}(c) shows a generalization of this effect 
which involves sites more distant from the impurity (hereafter
referred to as 
``second-rank Henley effect'') and which we find to be relevant for CuO.
If the wave-function at atom 1 is perturbed due to the impurity, all
the intersublattice interactions indicated by thick bonds get
affected. Weiss fields parallel to the odd spin magnetization  appear
also in all second neighbors of the impurity (defined as the first
neighbors of the first neighbors excluding the impurity, e.g. atom 2)
leading again to the stabilization of AF2. 

In order to explore the relevance of quenched disorder effects on CuO
we first use first-principle computations to derive a generalized
Heisenberg model of the magnetic degrees of freedom
(Sec.~\ref{sec:model-hamiltonian}) and use classical Monte Carlo
computations to show  that it can describe correctly the undoped phase
(Sec.~\ref{sec:finite-temp-simul}) including the subtle incommensurate
spiral. This is supplemented by an analytical computation of the pitch
of the spiral (Appendix~\ref{sec:yablonskii}). 
Then we use again first principle
computations to study systematically the effects of different
impurities on stabilizing the multiferroic phase
(Sec.~\ref{sec:dft-calculations}). Monte Carlo simulations confirm
that indeed impurities stabilize the ferroelectric phase respect to
the collinear magnetic phase
(Sec.~\ref{sec:finite-temp-simul-1}). Finally experiments using Zn and
Co as dopants confirm the theory prediction (Sec.~\ref{sec:experiments}).   
We conclude in Sec.~\ref{sec:conclusions}. 


\section{Undoped C\lowercase{u}O}
\label{sec:undoped-cuo}
Before approaching the doped compound, an accurate description of the undoped compound was desired as a reference. Therefore a parametrization to a magnetic Hamiltonian in the undoped phase has been performed which is treated semiclassically at finite temperatures in Monte Carlo simulations. This extends previous computations \cite{Filippetti05,Rocquefelte10,Giovannetti11,Jin12,Rocquefelte12} which considered a more limited set of exchange constants $J_{ij}$ acting among spins at sites $i$ and $j$. In addition our semiclassical model includes a biquadratic pairwise interaction $K$ between nearest neighbor spins. 

\subsection{Model Hamiltonian}
\label{sec:model-hamiltonian}
Total energies were calculated with density functional theory (DFT)
for 39 collinear spin 
configurations, from which a magnetic Hamiltonian,  
\begin{eqnarray}
&& \mathscr{H}_{\mathrm{M}} = \mathscr{H}_{\mathrm{exch}} +
\mathscr{H}_{\mathrm{bq}} + \mathscr{H}_{\mathrm{ani}} \\
&=&\frac{1}{2}\sum_{i\neq j}J_{ij}\mathbf{S}_i\cdot\mathbf{S}_j +
\frac{1}{2}\sum_{i\neq j}K_{ij}(\mathbf{S}_i\cdot\mathbf{S}_j)^2 +
\frac{1}{2}\sum_{i\neq j} \mathbf{S}_i \mathbf{J}_{ij}^{\mathrm{ani}}
\mathbf{S}_j \nonumber
\end{eqnarray}
was parametrized and expressed in terms of classical spin variables
$\mathbf{S}_i$ of length $1/2$. 

For the DFT calculations we have used the VASP software
\cite{Kresse96,Anisimov97} with a GGA+U functional \cite{Dudarev98} in
the PBE parametrization \cite{Perdew92,Perdew96}. The calculations
were performed for a 64 atom cell with the lattice vectors
$\mathbf{a}' =2 \mathbf{a}$, $\mathbf{b}' =2 \mathbf{b}$, $\mathbf{c}'
=2 \mathbf{c}$ with the experimental \cite{Asbrink70} atomic positions
and lattice vectors
$\mathbf{a}=(4.6837,0,0)$, $\mathbf{b}=(0,3.4226, 0)$ and $\mathbf{c}
= (0.85005, 0,5.0579)$~\AA. 
Convergence was obtained for a cutoff of 400 eV for the augmented
plane wave basis set and a 2 3 2 $\Gamma$-centered $k$ point mesh. For
the Cu atoms the Hubbard constant $U_{\mathrm{eff}}=5.5$~eV
\cite{Dudarev98} was used. This resulted in a magnetic moment of
$\mu_{Cu}=0.62$~$\mu_B$, corresponding well to the experimental value
of $0.65$~$\mu_B$ \cite{Forsyth88}. 

\begin{figure}
\includegraphics[width=0.50\textwidth]{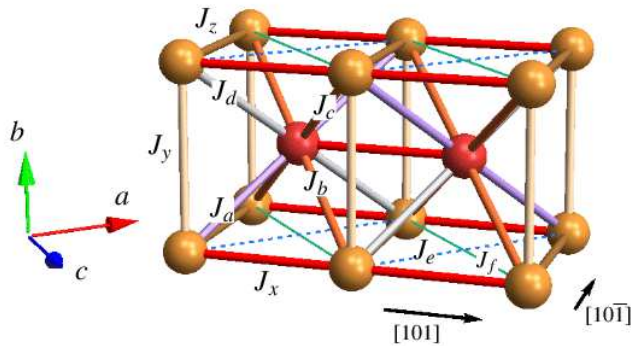}
\caption{\label{fig:MB_fig_3_9}The exchange couplings used in the calculations. Bonds of the same color are equivalent. Notice that the orientation of the bonds bridged by $J_a$ (violet) and $J_d$ (gray) alternate on the [101] direction. The same alternation occurs in $[10\overline 1]$ and [010] directions. $J_b$ and $J_c$ instead are equivalent by symmetry.  The lattice defined by the Cu atoms resembles a body center cubic  structure with atoms on even planes (orange) at the vertices of the cuboids and atoms in odd planes at the center (red). Not shown in the picture are the second  neighbor couplings $J_{a2}$, $J_{b2}$, $J_{c2}$ and $J_{d2}$ which act respectively on the same direction of $J_{a}$, $J_{b}$, $J_{c}$ and $J_{d}$. i.e. $J_{a2}$ bridges the two even atoms at opposite vertices of the cuboid that are connected by  $J_{a}$ with the odd atom (red) at the center. $J_{a2}$ and $J_{d2}$  follow the same alternation as $J_{a}$ and $J_{d}$. 
}
\end{figure}

A set of 10 effective $J_{ij}$ ranging up to a distance of
6.166~\AA~were considered (See Table~\ref{tab:exchcoup} and
Fig.~\ref{fig:MB_fig_3_9}). Note that positive numbers refer to an
antiferromagnetic exchange interaction. The total energies of the 39
spin configuration where computed not
including spin orbit coupling (SOC), rendering an overdetermined
equation system for 
the $J_{ij}$. For each configuration we defined the error as the
difference in energy between the result of the model and the DFT
computation. An optimum solution was obtained minimizing the least
mean-square error with respect to the exchange couplings
\cite{Rocquefelte11,Giovannetti11-1}. 

\begin{table*}
\caption{\label{tab:exchcoup}Heisenberg exchange couplings. Positive numbers refer to an antiferromagnetic exchange interaction. By symmetry $J_b=J_c$ and $J_{b2}=J_{c2}$. Analyzing the exchange paths we assume $J_{d2}=0$.}
\begin{ruledtabular}
\small
\begin{tabular}{|c|c|c|c|c|c|c|c|c|c|c|}
\hline
Label          & $J_a$ &  $J_c$ & $J_d$     & $J_x$   & $J_y$ & $J_z$ & $J_{e}$ & $J_{f}$ & $J_{a2}$ & $J_{c2}$ \\
Dist (\AA)     & 2.90 & 3.08 & 2.90 & 3.17 & 3.42 & 3.75 & 4.68 & 5.12 & 5.80 & 6.16 \\
Strength (meV) & 5.83 & -2.04 &  5.83     & -4.24   & -1.90 & 120.00& -2.61 & 16.10 & 24.30 & -1.77\\
\hline
\end{tabular}
\end{ruledtabular}
\end{table*}

The strongest coupling is  $J_z$ (c.f. Fig.~\ref{fig:MB_fig_3_9}) and
stabilizes the AFM chains running along $[10\bar 1]$  direction. The large
strength is explained by the Cu-O-Cu bond angle of $146^{\circ}$ close
to $180^{\circ}$. 
For $J_x$ the corresponding bond angle is $109^{\circ}$. The
closeness  to $90^{\circ}$ leads to a FM coupling according to
Goodenough-Kanamori rules \cite{Goodenough55,Kanamori63}. Along the
$y$-direction a coupling $J_y$ was considered to account for the
effective exchange between Cu atoms on planes separated by $\Delta
y=\pm 1$, i.e. a coupling between neighboring even planes or
neighboring odd planes.

The four couplings with the shortest distances link the even and odd
$y$-planes.  The linked atoms are defined ``nearest neighbors'' in
Fig.~\ref{fig:atoms}.  Even though the $J_a$ and $J_d$ couplings have the same
distance, close examination of the structure shows that they can have
different values. Despite that, in order to simplify the computations,
we assumed the same value $J_a=J_d$. An analytic computation shows
that this approximation does not affect the pitch of the spiral (see
Appendix~\ref{sec:yablonskii}).

Within constant $y$-planes the
exchange couplings $J_x$, $J_z$, $J_{e}$, and $J_{f}$ were
considered.  Going two steps along $J_a$, $J_b$, $J_c$ and
$J_d$ the couplings $J_{a2}$, $J_{b2}$, $J_{c2}$, $J_{d2}$ are
introduced with $J_{b2}=J_{c2}$ by symmetry. $J_{d2}$ was assumed to
be zero after a close examination of potential exchange paths.  


The effective biquadratic $K_{ij}=-K$ interaction acts among classical
neighboring spins on different sublattice in the spirit of
Ref.~\cite{Nikuni98} and takes into account charge relaxation
effects. Such interaction, which for spin-half systems has not an
obvious quantum counterpart \cite{Nikuni98}, describes the effect of
relaxation of charge degrees of freedom in stabilizing the AF1
configuration with respect to the AF2 phase. It was determined by
rotating the spins on the even plane to be perpendicular to the spins
on the odd plane. For simplicity we restrict the biquadratic
interaction to the 8 bonds connecting spins on different sublattices
(the bonds bridged by $J_a$, $J_b$, $J_c$ and $J_d$) and assume all
the bonds have the same constant $K_{ij}=-K$.

 When allowing the charge
to relax in the DFT calculations we find that at zero doping the  AF2
phase is higher in energy with respect to the AF1 phase by $\Delta
E(0)=2.15$ meV/Cu. The energy splitting is parametrized as $\Delta
E(0)=4KS^4$  with $S=1/2$, for the Cu spin and $K=8.61$~meV.  

Contributions to symmetric anisotropic exchange was determined by including SOC in the calculation and from Ref.~\cite{Villarreal12}.  We assign anisotropic interactions to the strongest magnetic bond so that ${\bf J}_{ij}^{\rm ani}\ne 0$ only for bonds bridged by $J_z$. Inclusion of  spin-orbit interaction in the calculation reveals that the $y$ direction ([010]) acts as an easy axis with  ${\bf J}_{ij}^{\rm ani}|_{yy}=0.145$~meV. In order to reproduce correctly the orientation of the spins in the AF2 phase a more precise description of anisotropic exchange is necessary. Following \cite{Villarreal12} we use ${\bf J}_{ij}^{\rm ani}|_{zz} = 0.5 {\bf J}_{ij}^{\rm ani}|_{yy}$ and ${\bf J}_{ij}^{\rm ani}|_{xz} = 0.42 {\bf J}_{ij}^{\rm ani}|_{zz}$ with the Cartesian $x$-axis coinciding with the [100] direction. 

\subsection{Finite temperature simulations}
\label{sec:finite-temp-simul}
We have performed Monte Carlo Metropolis (MC) simulations of the above Hamiltonian. To improve the convergence and determine accurately the transitions temperatures the parallel tempering scheme \cite{Hukushima96} was used. Cell size for the phase diagram was chosen not too large to avoid the exponential critical slowing down typical of first-order phase transitions (see Ref.~\cite{Martin-Mayor07} and Appendix~\ref{sec:appptmc}). The magnetic ordering has been studied by computing the static structure factor $S(\mathbf{q})$, calculated as the Fourier transform of the spatial displaced equal time spin-spin correlation function
\begin{eqnarray}
S(\mathbf{q})=\frac{1}{2\pi}\int d\mathbf{r} \;
e^{i\mathbf{q}\cdot(\mathbf{r}-\mathbf{r}')} \langle\mathbf{S(r)}\cdot\mathbf{S(r')}\rangle.
\end{eqnarray}
Here $\langle\ldots \rangle$ indicate an average over the $N$ Cu sites
in the simulation cell and over time.

The ferroelectric properties of the system were investigated by sampling the average polarization \cite{Katsura05} 
\begin{eqnarray}
\mathbf{P}=\frac{\gamma}{N}\sum_{\langle i j \rangle}\hat{\mathbf{e}}_{ij}\times(\mathbf{S}_i\times\mathbf{S}_j)
\end{eqnarray}
with the sum running over the first neighbors defined in Fig.~\ref{fig:atoms}. $\hat{\mathbf{e}}_{ij}$ is a vector in the direction of the bond and $\gamma=894$ $\mu$C/m$^2$ has been determined by matching the electronic DFT polarization in the AF2 state at $T=0$~K and without impurities \cite{Giovannetti11}. 

\subsection{Results}

 We use the peak value of the structure factor  $S(\mathbf{q})$ as a function of momentum as the square of the order parameter for the two magnetic phases. The corresponding momentum is commensurate for the AF1 phase, $\mathbf{q}_{AF1}=(0.5, 0, -0.5)$, and incommensurate for the AF2 phase, $\mathbf{q}_{AF2}=(0.528, 0, -0.472)$ in excellent agreement with the experimental \cite{Forsyth88,Yang89,Brown91} value $\mathbf{q}_{AF2}=(0.506, 0, -0.483)$. A close by value can be obtained analytically providing a stringent test for the model (See Appendix~\ref{sec:yablonskii}).

In Fig.~\ref{fig:PhaseDiagramCuO} we see that a first order transition occurs at $T_{N1}=165$~K. Above this temperature and below $T_{N2}=178$~K (determined from the anomaly in the specific heat $C_V$) the incommensurate AF2 phase appears. Above $T_{N2}$ the system becomes paramagnetic. In Refs.~\cite{Giovannetti11,Jin12} an electron-phonon coupling was deemed necessary to stabilize the AF2 state. With our present, more accurate parameter set and MC simulation, we find that AF2 is stable even without coupling to the lattice, in agreement with Ref.~\cite{Villarreal12,Yablonskii90}. In addition, the temperature dependence of the structure factor is in good agreement with neutron scattering experiments \cite{Chatterji05}. 

\begin{figure}
\includegraphics[width=0.50\textwidth]{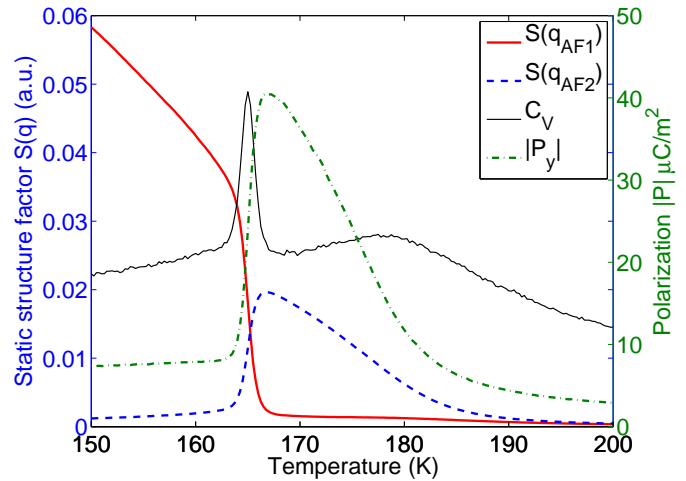}
\caption{\label{fig:PhaseDiagramCuO}Order parameters for undoped CuO from classical Monte Carlo simulations. The static structure factor $S(\mathbf{q})$ for $\mathbf{q}_{AF1}$ and $\mathbf{q}_{AF2}$, and the out of plane polarization $P_y$, for a $36\times 4\times 36$ cell. Simulation cell sizes are stated in terms of repetitions $L_a \times L_b \times L_c$ along the lattice vectors $\mathbf{a},\mathbf{b},\mathbf{c}$ of the conventional unit cell with 4 Cu atoms. 
}  
\end{figure}

Between $T_{N1}$ and  $T_{N2}$ also the chiral symmetry is
spontaneously broken and the system acquires a net polarization due to
the inverse Dzyaloshinskii-Moriya mechanism \cite{Katsura05} as shown
in Fig.~\ref{fig:PhaseDiagramCuO}. While in principle it is possible
that the polarization becomes finite above the temperature at which
the magnetic ordering sets in, we do not find evidence of such
``chiral liquid'' phase or the AF3 phase of Ref.~\cite{Villarreal12},
although we can not exclude them due to size limitations.

Overall the magnetic phase diagram is in good accord with
experiment. We remark that we have not made any fitting of the
exchange constants to reproduce the experimental N\'eel
temperatures. Probably our exchange constants are underestimated
leading to both $T_{N1}$ and $T_{N2}$ being 23\% lower than the
experimental values. Qualitatively similar results were obtained using
the parameters of Refs.~\cite{Giovannetti11} and \cite{Rocquefelte10}
except that they have stronger frustration leading to spirals with a
shorter periodicity.

\section{Doped C\lowercase{u}O}
\label{sec:doped-cuo}
\subsection{DFT calculations}
\label{sec:dft-calculations}
\begin{figure}
\includegraphics[width=0.50\textwidth]{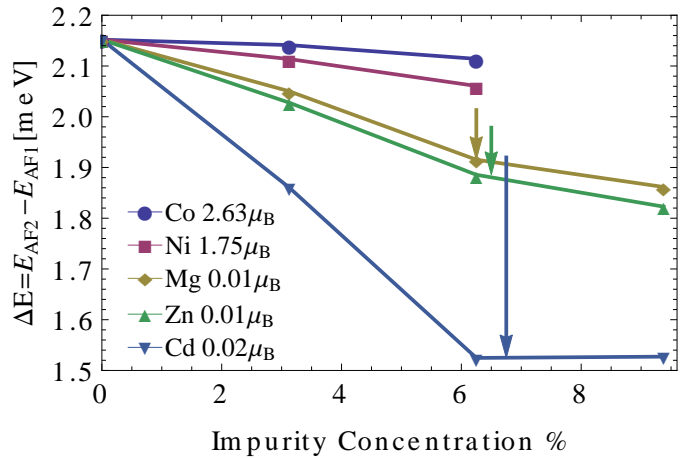}
\caption{\label{fig:DFTdeltaE}DFT computation of the stabilization of the ferroelectric phase with different impurities. We show the energy difference per Cu atom between the AF1 and the AF2 configurations as a function of doping for different dopants. 
The arrows show the magnitude of Henley relaxation energy at 6.250\% doping for (from left to right) Mg, Zn and Cd. The legend reports the magnetic moment we find at the impurity site in the DFT computations.
}
\end{figure}

As mentioned in Sec.~\ref{sec:intro}, doping is expected to reduce the
energy gap between AF2 and AF1. The magnitude of this effect depends
on details such as how much the local interactions are modified by the
impurities, their magnetic moment, etc. To estimate these effects we
use DFT computations to compute the energy gain of the AF2 phase
respect to the AF1 phase due to different dopants. For simplicity we
neglect the small incommensurability and spin-orbit coupling. These
effects will be restored below. 
We have replaced up to 3 out of
the 32 Cu atoms in the cell with impurities. In the case of magnetic
impurities Hubbard constants have been used, with values taken from
the literature; for Co, $U_{\mathrm{eff}}=3.3$~eV \cite{Wang06-1}, and
for Ni, $U_{\mathrm{eff}}=7.05$~eV \cite{Dudarev98}. The resulting
magnetic moments \cite{note} reported on the inset of
Fig.~\ref{fig:DFTdeltaE} are  in good agreement with published
values \cite{Rohrbach04,Chen11-1}.  
 
As expected, we find that for each of the impurity elements the energy
difference per Cu among the two states, $\Delta E$, decreases
monotonically with the impurity concentration
(c.f. Fig.~\ref{fig:DFTdeltaE}). Up to $\sim $6\% doping the decrease
in the energy gap is linear.  For higher doping the non-linear
behavior indicates that impurities start to interact with each other.


In order to understand the energy gain it is useful to separate the
energy difference between the two phases in two contributions 
\begin{equation}
  \label{eq:deltae}
\Delta  E=\Delta E_{\rm bq}+\Delta E_{\rm Hen}.
\end{equation}
The first term is due to the local change of the
interactions which contribute to $\Delta E$ in the undoped case. This
can be obtained by computing the energy in a configuration in which
different sublattice spins in the AF2 phase are constrained to be exactly at right
angles. In this case the contribution of classical Heisenberg
interactions cancels and only biquadratic and spin-orbit terms
contribute.  We checked that the latter makes a negligibly
doping-dependent contribution (on the order of 2$\mu eV$ at 3\%
doping) therefore this part of the energy is referred to as ``dilution
in the biquadratic interaction'' ($\Delta E_{\rm bq}$). 

In the case of non-magnetic impurities and assuming that the biquadratic interactions do not change
around the impurities one obtains  $\Delta E_{\rm
  bq}(x)=\Delta E(0)(1-2x)$. We find that this formula overestimates the DFT doping
induced dilution energy gain for Zn and Mg and works much better for
Cd. We attribute this discrepancy to local rearrangements of the
biquadratic interactions on the sites neighboring the
impurity. 

The second term in Eq.~\eqref{eq:deltae} is Henley relaxation energy
($\Delta E_{\rm Hen}$) which is the energy gain obtained from the
previous configurations allowing the spins to relax in the transverse
direction.   The DFT Henley relaxation energy is shown at 6.250\%
doping by the arrows in Fig.~\ref{fig:DFTdeltaE} and accounts for
around 40\% of the gap reduction for Zn and Mg, while it dominates the
doping induced energy gain for Cd.  As expected, the transverse relaxation is
negligible in the AF1 configuration so the effect reported is due to
the transverse relaxation of the magnetization in the  AF2
configuration.

From a rough estimate, the conventional Henley \cite{Henley89}
mechanism  which we call ``first-rank" [Fig.~\ref{fig:atoms}(b)],
leads to an energy gain of order $\Delta E_{\rm Hen}\sim - x
J_1^2/J_2~\sim- x\; 0.09$~meV with $J_1\sim 3$~meV of the order of the
intersublattice exchanges and $J_2\sim 100$ meV of the order of the
intrasublattice exchange. This is much smaller than $\Delta
E_{bq}$. The sizable Henley relaxation found is due to higher-rank
Henley effect. Figure~\ref{fig:atoms}(c) shows schematically an example of
second-rank Henley effect which comes into play do to the disturbance 
of the spin and charge density beyond the first shell of neighbors of
the impurity. For the atom labeled 2 the Heisenberg interactions on
the orange links become unbalanced, thus a local horizontal Weiss
field appears which stabilizes AF2. Analyzing the spin configurations, we find
that higher-rank effects are particularly important for Cd consistent
with the larger Henley relaxation energy.   

\subsection{Finite temperature simulations}
\label{sec:finite-temp-simul-1}
Since non-magnetic impurities show the largest stabilization effect of
the AF2 phase, we concentrate on those. It is expected that the
ordering temperature will be determined mainly by the intrasublattice
interactions so $T_{N2}$ will be depressed by dilution. On the other hand, once the
system orders, the width in temperature of the multiferroic 
phase will be determined by the relative stability among the phases. 

\begin{figure}
\includegraphics[width=0.50\textwidth]{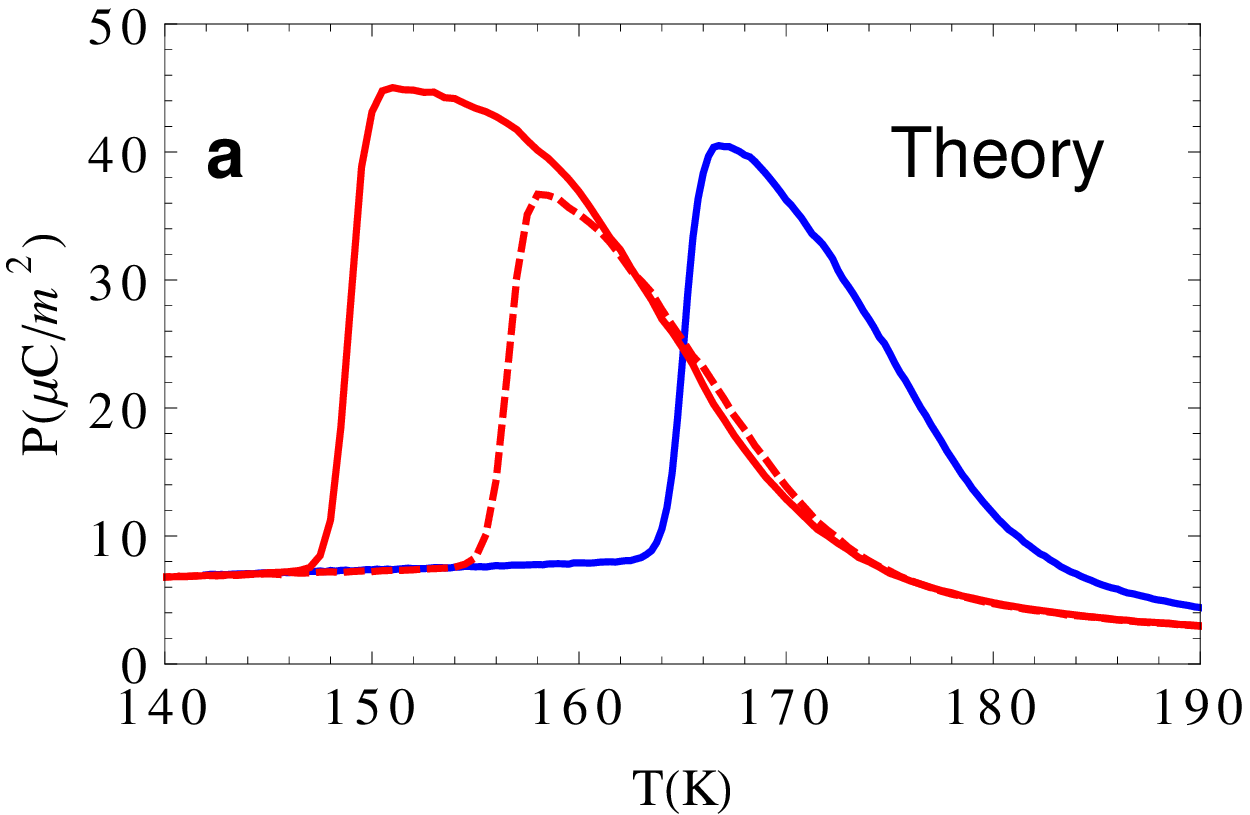}
\includegraphics[width=0.50\textwidth]{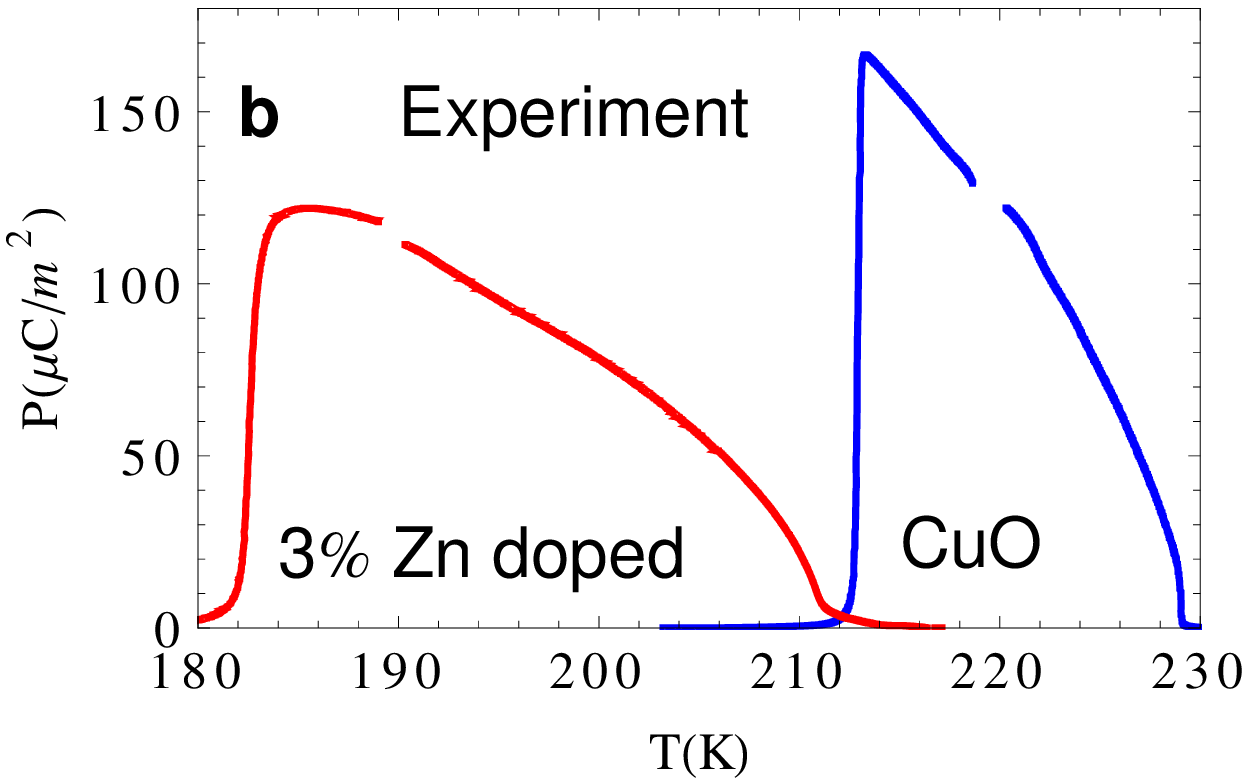}
\caption{\label{fig:PhaseDiagramDopedCuO}$b$-axis electric polarization as a function of temperature. (a) Monte Carlo computations for undoped CuO (blue) and for 3.125\% non-magnetic doping  without enhancement of interactions (dashed, red line) and with an enhancement factor $\eta=3$ (full red line). (b) Experiment using the same protocol as in Ref.~\cite{Kimura08}.
}  
\end{figure}

 First we analyze first-rank Henley effect alone. For this we simply
 canceled all interactions connecting the impurity site 
to the rest of the system.  Fig.~\ref{fig:PhaseDiagramDopedCuO}(a)
shows the polarization from MC simulations as a function of
temperature in the doped (dashed line) and undoped case (blue
line). In this case we find that $T_{N1}$ and $T_{N2}$ decreases by
the same amount so the temperature width of the AF2 phase remains the
same. The decrease in the magnitude of the polarization is due to
incoherent canting effects induced by the impurity and the shift to
lower temperatures is the expected effect of dilution on the intrasublattice
interactions.

Since the first-rank Henley relaxation energy is very small in this
system, the fact that the temperature width of the AF1 phase remains
constant means that the relative stability among the two phases is not
affected by simply dilution effects. Indeed we have mentioned above
that dilution penalizes the AF1 energy by a $(1-2x)$ factor. The
stability of the AF2 phase is due to Yablonskii's mechanism (Appendix~\ref{sec:yablonskii} and Ref.\cite{Yablonskii90}), and it is
penalized by the same factor, which is a 
consequence of the pairwise nature of interactions. Thus the internal
energy of both phases is penalized in the same way which leads to a
crossing of the free energies at a doping-independent temperature
distance from the instability of the paramagnet (roughly
$T_{N2}$). This  delicate balance can be altered  by the relaxation
energies due to higher-rank Henley effects. To demonstrate this we
multiply all four inter-sublattice exchange couplings on the eight Cu
atoms that are nearest neighbors to an impurity by the same factor
$\eta$. These interactions are very small ($\pm$ a few meV) because
the exchange paths form an angle close to 90$^o$. As in the case
\cite{Braden96} of CuGeO$_3$ we expect that even weak perturbations of
the electronic orbital overlaps (here due to the presence of the
impurities) can result in large relative changes of these
intersublattice couplings.  Henley effects coming from more distant
sites  are incorporated for simplicity in an effective way in the
single $\eta$ parameter.  
For $\eta=3$ one obtains a stabilization of the AF2 due to Henley mechanism comparable with DFT for Zn (i.e. -15 $\mu$eV to be compared with -32 $\mu$eV in DFT for 3.125\% doped Zn).
In Fig.~\ref{fig:PhaseDiagramDopedCuO}(a) we show the polarization for the doped $\eta=3$ case (red full line). Even if the extra stabilization energy of AF2 is modest, a large increase on the width of the AF2 phase is observed.  

\subsection{Experiments}
\label{sec:experiments}

In order to check the theory experiments have been perform.
Single crystals of undoped, Zn-doped, and Co-doped CuO were grown
under about 8 atm of pure oxygen by the floating zone technique,
following Ref.~\cite{Prabhakaran03}. The grown crystals were oriented
using X-ray diffractometers, and cut into thin plates with the widest
faces perpendicular to the $b$ axis. 

For the measurements of
pyroelectric current, silver electrodes were vacuum-deposited onto the
widest faces of the plate-shaped crystals. The electric polarization
along the $b$ axis was obtained from integration of the pyroelectric
current over time. Before the respective measurements, poling electric
fields were applied to the crystals at the paramagnetic phase. Then
the crystals were cooled to the lower boundary of the AF2 phase. After
these procedures, the poling electric fields were removed, and the
pyroelectric current was measured during the temperature up and down
sweeps as in Ref.~\cite{Kimura08}.

The main experimental result is that the multiferroic window indeed
widens with doping. The temperature dependence of the
polarization for the doped and undoped compound 
[Fig.~\ref{fig:PhaseDiagramDopedCuO}(b)] are in good agreement with
the theory [Fig.~\ref{fig:PhaseDiagramDopedCuO}(a)]. To the best of
our knowledge this is the first verification of Henley's quenched
disorder mechanism at work.

\begin{figure}
\includegraphics[width=0.50\textwidth]{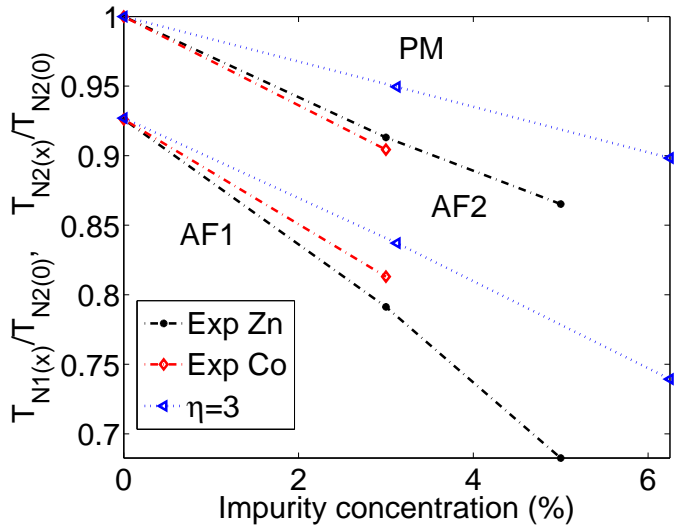}
\caption{\label{fig:phasediag}Experimental and theoretical phase diagram as a function of doping. We plot the N\'eel  temperatures as a function of doping for the experiment with Zn (dots) and Co (diamonds) together with the theoretical results for $\eta=3$ (triangles).
}  
\end{figure}

Fig.~\ref{fig:phasediag} shows a comparison of the theoretical and experimental phase diagram as a function of doping. As noted above $T_{N2}$ is reduced with doping and this effect is underestimated in the computations, probably due to the lack of quantum fluctuations which will tend to suppress the ordering temperature. More important for our present focus, at around 5\% Zn doping \footnote{Previous results for the magnetic anomalies at 5\% Zn doping are in well agreement for $T_{\textrm{N2}}$ but yield a smaller widening of the window ($T_{\textrm{N1(5\%)}}/T_{\textrm{N2,(0)}}=0.74$) \cite{Prabhakaran03}. Reexamination of these samples showed that they where weakly conducting as opposed to the samples used in Fig.~\ref{fig:phasediag} which where insulating.} the experimental multiferroic window widens by 250\% which should be compared with the theory for $\eta=3$  which yields a 200\% widening. Notice that the width of the AF2 region has been found to increase with positive pressure \cite{Chatterji05,Rocquefelte12,Rocquefelte13}; however, chemical pressure can not explain the effect. Indeed since Zn is larger than Cu the effect of Zn doping will correspond to a negative pressure \footnote{We have checked with powder x-ray diffraction that the volume does not decrease with Zn doping.} which will yield the opposite effect. Interestingly in the case of Mg which is smaller than Cu it is likely that the two effects become additive with the extra bonus of an increase (or a smaller reduction) in the N\'eel temperatures. Experiments and theory (with the addition of lattice relaxation effects which were here neglected) are underway to consider this promising case. We also see that experimentally the effect with Co is smaller than with Zn which is in qualitative agreement with the smaller stabilization of the AF2 phase for Co found in Fig.~\ref{fig:DFTdeltaE}. We predict the AF2 stabilization effect to be quite large for Cd due to the sizable higher-rank Henley effect found.  

\section{Conclusions}
\label{sec:conclusions}
We have shown theoretically  and experimentally  that disorder in the
form of impurities in CuO can stabilize the ferroelectric phase with
respect to the non-ferroelectric collinear magnetic phase. CuO consists of two sublattices with strong intrasublattice interactions and weak inter-sublattice interactions arranged in such a way that the classical Weiss field of one sublattice on the other cancel. This allows intrasublattice ordering at high temperatures leaving at the same time the possibility to manipulate the relative orientation of the sublattice spins with weak perturbations like a small amount of impurities. This peculiar symmetry and separation of energy scales allows ferroelectricity induced by magnetism at high temperatures as opposed to other frustrated systems which are multiferroic only at very low temperatures \cite{Cheong07}. Unfortunately impurities affect the intrasublattice interactions which has the unwanted feature to reduce $T_{N2}$. As mentioned above, in the case of CuO this may be counteracted by chemical pressure.
  
On a wider perspective we notice that the cancellation of Weiss field
is not unique of CuO but can be found in other systems as well
\cite{Gukasov88,Giebultowicz93,Capriotti04,2005FrustratedDiep,2011FrustratedLacroix,Savary12}. Since
negative biquadratic terms are the rule (even without invoking quantum
and thermal order-by-disorder mechanisms \cite{Henley89}), collinear
ground states are the most probable outcome. However by the same
mechanism shown here, impurities can transform an unsuspected
collinear antiferromagnet   (possibly using strains and layering to
modify the symmetry) into  a multiferroic, providing a novel route to
search and engineer multiferroics. Particularly interesting in this
regard are fcc antiferromagnets \cite{Henley89} which also have a
quasidegenerate classical ground state as for example MnTe
\cite{Giebultowicz93} or even NiO which has a pristine N\'eel temperature of 525~K providing a good starting point for engineering of a room temperature multiferroic. 

\begin{acknowledgments}
J. L. is supported by the Italian Institute of Technology through
seed project NEWDFESCM. J. L. acknowledges hospitality by the Aspen
Center for Physics under National Science Foundation Grant
No. PHYS-1066293. S. P. is supported by the European Research 
Council (ERC) grant 203523 (BISMUTH). O. E. is supported from ERC
grant 247062 (ASD) and the KAW foundation. A. B. receives funding
from eSSENCE. L. B. receives funding from SeRC and the G\"{o}ran
Gustafsson Foundation. A. B., L. B., O. E., and J. H. are supported
by the Swedish Research Council. T.U. and T.K are supported by KAKENHI
(Grant No. 24244058), MEXT, Japan. We acknowledge computational
resources and support from CASPUR, CINECA, UoS Napoli CNR-SPIN, and
from SNIC. 
\end{acknowledgments}


\appendix
\section{Stabilization of the AF2 phase by Yablonskii mechanism}\label{sec:yablonskii}
As discussed in Sec.~\ref{sec:intro} the magnetic structure of CuO can be divided in two sublattices. Within each sublattice spins orient antiferromagnetically in the $[10\bar 1]$ direction and ferromagnetically in the $y$ and $[101]$ directions. Neglecting biquadratic and anisotropic interactions if the spins are collinear within one sublattice, the magnetic ground state is degenerate at the classical level. In this Appendix we show how this degeneracy can be broken by allowing a long wave-length spiral \cite{Yablonskii90}.

\begin{figure}
\includegraphics[width=0.50\textwidth]{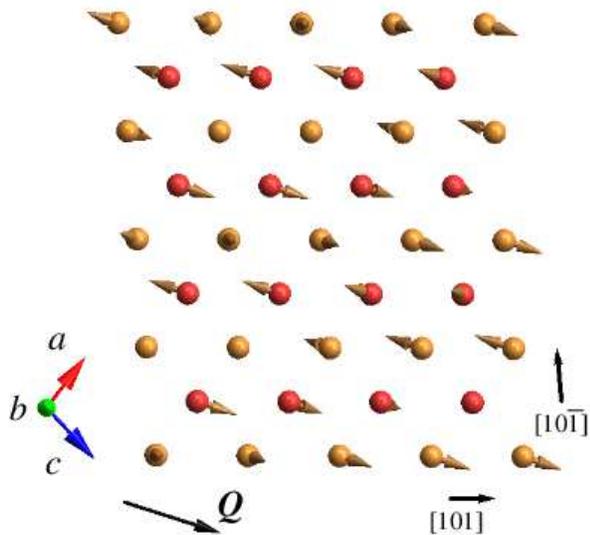}
\caption{\label{fig:spiral}The incommensurate AF2 configuration.  Cu ions in an even (odd) plane are represented by orange (red) spheres. Brown arrows represent spins. The pitch of the spiral has been increased by a factor of 40 to allow the visualization of the spin rotations. $\bf Q$ is the ordering wave vector of the spiral distortion.
} 
\end{figure}

Fig.~\ref{fig:spiral} shows schematically the incommensurate structure present in the AF2 configuration. The experimental periodicity is a factor of 40 longer than what represented. It is easy to check that in the presence of a spiral the spins in the even sublattice produce a Weiss field in the odd sublattice which is perpendicular to the local spin direction in the even sublattice (and vice versa). Thus the long-wave length spiral favors a configuration where locally the spins are nearly perpendicular on different sublattices i.e. the AF2 configuration. 

The plane of the spiral is determined by the anisotropies in the system and is defined by the ${\bf b}$ axis and a vector ${\bf v}$ in the (010) plane. ${\bf v}$ is nearly parallel to the vector ${\bf Q}$ defined below and shown in Fig.~\ref{fig:spiral}. For the present parameter set the direction of  ${\bf v}$ agrees with the experimental one (see below). For the moment we set $T=0$~K and neglect the anisotropies taking the direction of ${\bf v}$ as granted. The spin texture is given by, 
\begin{equation}
{\bf S}({\bf r})=S [ \hat{\bf b} \sin(2\pi {\bf q}_{\rm AF2} \cdot{\bf r})+\hat{\bf v}\cos(2\pi {\bf q}_{\rm AF2}\cdot{\bf r})]
\end{equation}
with ${\bf q}_{\rm AF2}={\bf q}_{\rm AF1}+{\bf Q}$. The vector ${\bf Q}$ is shown schematically in Fig.~\ref{fig:spiral}. Notice that the ordering vectors are also in the (010) plane and therefore ${\bf S}({\bf r})$ is constant in the ${\bf b}$ direction.  

The Weiss field acting on a spin in one sublattice due to the eight neighboring spins in the other sublattice is,
\begin{eqnarray}
{\bf h}_{1}({\bf r})=&-& 2\bar J_a \left[{\bf S}\left({\bf r}+\frac{{\bf
      a}+{\bf b}}2\right)+{\bf S}\left({\bf r}-\frac{{\bf a}+{\bf
      b}}2\right)\right]\nonumber\\
 &-&2J_c\left[{\bf S}\left({\bf r}+\frac{{\bf
      c}+{\bf b}}2\right)+{\bf S}\left({\bf r}-\frac{{\bf c}+{\bf
      b}}2\right)\right]\nonumber\\
=&& 4 {\bf S}({\bf r})  \left[\bar J_a \sin(\pi {\bf Q}\cdot {\bf a})-J_c \sin(\pi {\bf Q}\cdot {\bf c})\right]\nonumber\\
\end{eqnarray}
where we used the translational invariance of the structure along ${\bf b}$ and defined $\bar J_a=(J_a+J_d)/2$. For small ${\bf Q}$ the Weiss field is linear in ${\bf Q}$ producing a linear gain in the energy which renders a commensurate structure with the spins perpendicular among different sublattices unstable to spiral formation. It is convenient to parametrize the ordering wave vector in terms of reciprocal lattice vectors, ${\bf Q}=\delta_a \mathbf{a}^*+ \delta_c \mathbf{c}^*=(\delta_a,0,\delta_c)$~r.l.u. Neglecting the anisotropic contributions but restoring the biquadratic interaction the total energy per Cu ion reads, 
\begin{eqnarray}
\frac{E_{\rm AF2}({\bf Q})}{S^2} &=& -2\bar{J}_a\sin\left(\pi\delta_a \right) + 
2J_c\sin\left(\pi\delta_c \right) + J_y \nonumber\\
&& +J_x\cos\left[\pi( \delta_a+\delta_c) \right]- J_z \cos\left[\pi( \delta_a-\delta_c) \right] \nonumber\\ 
&& -(J_e+J_{a2}+J_{d2})\cos\left(2\pi \delta_a \right) \nonumber\\
&& -(J_f+2J_{c2})\cos\left(2\pi \delta_c \right)\nonumber\\
&& -2KS^2\left[\sin^2\left(\pi\delta_a \right) + \sin^2\left(\pi \delta_c \right)\right].
\end{eqnarray}
Minimization with respect to the ordering wave-vector can be easily done expanding the energy to second order in ${\bf Q}$. One finds ${\bf q}_{\rm AF2}=(0.538, 0, -0.466)$ at $T=0$~K for the present parameter set. We have also done a few Monte Carlo runs in cells of size $200 \times 4 \times 200 $ to increase the momentum resolution yelding ${\bf q}_{\rm AF2}=(0.530, 0,-0.475)$ for $T_{\rm N1}<T<T_{\rm N2}$. Both results are close to the experimental wave-vector \cite{Forsyth88,Yang89,Brown91} and to the incomensurability in the cells used to determine the whole phase diagram. 

Taking the commensurate AF2 state as a reference the energy of the AF2 state is, 
\begin{equation}
E_{\rm AF2}({\bf Q})-E_{\rm AF2}({\bf Q}=0)=-0.23 \;{\rm  meV},
\end{equation}
to be compared with the energy of the AF1 state, 
\begin{equation}
E_{\rm AF1}-E_{\rm AF2}({\bf Q}=0)=-2.15 \; {\rm meV}.
\end{equation}
Thus one finds that at $T=0$~K the AF1 state is the ground state, as found experimentally. Notice that the above energies scale as $S^2$ for the AF2 state and as $S^4$ for the AF1. At finite temperatures these energies are suppressed in modulus by factors of the form $\langle\mathbf{S}_i\cdot\mathbf{S}_j\rangle$ with $i$, $j$ nearest neighbors for the AF2 state and $\langle(\mathbf{S}_i\cdot\mathbf{S}_j)^2\rangle$ for the AF1 state. In a simple mean-field picture $\langle\mathbf{S}_i\cdot\mathbf{S}_j\rangle$ vanishes as a power law as a function of temperature at $T_{\rm MF}\sim T_{N2}$ while $\langle(\mathbf{S}_i\cdot\mathbf{S}_j)^2\rangle$ vanishes faster. Thus below $T_{\rm MF}$ there exist a temperature window where the AF2 phase is more stable than the AF1 phase as found experimentally and in our numerical simulations \cite{Yablonskii90,Johnson12,Villarreal12}. We have verified in our numerical simulations that the spin configuration is planar, and the vector {\bf v} is at an angle of 70$^{\circ}$ to the $\mathbf{a}$-axis when $\mathscr{H}_{\mathrm{ani}}$ is included in good agreement with the experiment \cite{Forsyth88,Yang89,Brown91}.

\section{Parallel tempering Monte Carlo}\label{sec:appptmc}
The phase diagrams of the magnetic Hamiltonians has been worked out in classical Monte Carlo simulations combining a parallel tempering scheme \cite{Hukushima96} with a standard Metropolis local update algorithm. In a typical simulation a set of 64 replicas was used to cover a temperature range of 16 K. The temperatures were uniformly distributed with $\Delta T=0.25$~K. For the data presented in Figs.~\ref{fig:PhaseDiagramCuO} and \ref{fig:PhaseDiagramDopedCuO}, 
four non-overlapping temperature brackets were used, with the endpoint temperatures of the brackets chosen to be away from the phase transitions. Exchange of temperatures were attempted after each 100th Monte Carlo sweep (MCS), and restricted to pairs of neighboring temperatures. As the temperature brackets were rather narrow in comparison with the absolute temperature, only a weak increase in the acceptance rate could be seen for the upper temperatures in a bracket compared with the lower temperatures. For the simulation cells with size $36\times 4 \times 36$ unit cells, the acceptance rate were in the range 70\% to 80\%. The simulations were run over $5\cdot 10^6$ MCS, with the first $5\cdot 10^5$ MCS used
for equilibration.


\begin{thebibliography}{46}%
\makeatletter
\providecommand \@ifxundefined [1]{%
 \@ifx{#1\undefined}
}%
\providecommand \@ifnum [1]{%
 \ifnum #1\expandafter \@firstoftwo
 \else \expandafter \@secondoftwo
 \fi
}%
\providecommand \@ifx [1]{%
 \ifx #1\expandafter \@firstoftwo
 \else \expandafter \@secondoftwo
 \fi
}%
\providecommand \natexlab [1]{#1}%
\providecommand \enquote  [1]{``#1''}%
\providecommand \bibnamefont  [1]{#1}%
\providecommand \bibfnamefont [1]{#1}%
\providecommand \citenamefont [1]{#1}%
\providecommand \href@noop [0]{\@secondoftwo}%
\providecommand \href [0]{\begingroup \@sanitize@url \@href}%
\providecommand \@href[1]{\@@startlink{#1}\@@href}%
\providecommand \@@href[1]{\endgroup#1\@@endlink}%
\providecommand \@sanitize@url [0]{\catcode `\\12\catcode `\$12\catcode
  `\&12\catcode `\#12\catcode `\^12\catcode `\_12\catcode `\%12\relax}%
\providecommand \@@startlink[1]{}%
\providecommand \@@endlink[0]{}%
\providecommand \url  [0]{\begingroup\@sanitize@url \@url }%
\providecommand \@url [1]{\endgroup\@href {#1}{\urlprefix }}%
\providecommand \urlprefix  [0]{URL }%
\providecommand \Eprint [0]{\href }%
\providecommand \doibase [0]{http://dx.doi.org/}%
\providecommand \selectlanguage [0]{\@gobble}%
\providecommand \bibinfo  [0]{\@secondoftwo}%
\providecommand \bibfield  [0]{\@secondoftwo}%
\providecommand \translation [1]{[#1]}%
\providecommand \BibitemOpen [0]{}%
\providecommand \bibitemStop [0]{}%
\providecommand \bibitemNoStop [0]{.\EOS\space}%
\providecommand \EOS [0]{\spacefactor3000\relax}%
\providecommand \BibitemShut  [1]{\csname bibitem#1\endcsname}%
\let\auto@bib@innerbib\@empty
\bibitem [{\citenamefont {Wang}\ \emph {et~al.}(2009)\citenamefont {Wang},
  \citenamefont {Liu},\ and\ \citenamefont {Ren}}]{Wang09}%
  \BibitemOpen
  \bibfield  {author} {\bibinfo {author} {\bibfnamefont {K.}~\bibnamefont
  {Wang}}, \bibinfo {author} {\bibfnamefont {J.-M.}\ \bibnamefont {Liu}}, \
  and\ \bibinfo {author} {\bibfnamefont {Z.}~\bibnamefont {Ren}},\ }\href
  {\doibase 10.1080/00018730902920554} {\bibfield  {journal} {\bibinfo
  {journal} {Advances in Physics}\ }\textbf {\bibinfo {volume} {58}},\ \bibinfo
  {pages} {321} (\bibinfo {year} {2009})}\BibitemShut {NoStop}%
\bibitem [{\citenamefont {Kimura}\ \emph {et~al.}(2008)\citenamefont {Kimura},
  \citenamefont {Sekio}, \citenamefont {Nakamura}, \citenamefont {Siegrist},\
  and\ \citenamefont {Ramirez}}]{Kimura08}%
  \BibitemOpen
  \bibfield  {author} {\bibinfo {author} {\bibfnamefont {T.}~\bibnamefont
  {Kimura}}, \bibinfo {author} {\bibfnamefont {Y.}~\bibnamefont {Sekio}},
  \bibinfo {author} {\bibfnamefont {H.}~\bibnamefont {Nakamura}}, \bibinfo
  {author} {\bibfnamefont {T.}~\bibnamefont {Siegrist}}, \ and\ \bibinfo
  {author} {\bibfnamefont {A.~P.}\ \bibnamefont {Ramirez}},\ }\href {\doibase
  10.1038/nmat2125} {\bibfield  {journal} {\bibinfo  {journal} {Nature
  Materials}\ }\textbf {\bibinfo {volume} {7}},\ \bibinfo {pages} {291}
  (\bibinfo {year} {2008})}\BibitemShut {NoStop}%
\bibitem [{\citenamefont {Babkevich}\ \emph {et~al.}(2012)\citenamefont
  {Babkevich}, \citenamefont {Poole}, \citenamefont {Johnson}, \citenamefont
  {Roessli}, \citenamefont {Prabhakaran},\ and\ \citenamefont
  {Boothroyd}}]{Babkevich12}%
  \BibitemOpen
  \bibfield  {author} {\bibinfo {author} {\bibfnamefont {P.}~\bibnamefont
  {Babkevich}}, \bibinfo {author} {\bibfnamefont {A.}~\bibnamefont {Poole}},
  \bibinfo {author} {\bibfnamefont {R.~D.}\ \bibnamefont {Johnson}}, \bibinfo
  {author} {\bibfnamefont {B.}~\bibnamefont {Roessli}}, \bibinfo {author}
  {\bibfnamefont {D.}~\bibnamefont {Prabhakaran}}, \ and\ \bibinfo {author}
  {\bibfnamefont {A.~T.}\ \bibnamefont {Boothroyd}},\ }\href {\doibase
  10.1103/PhysRevB.85.134428} {\bibfield  {journal} {\bibinfo  {journal}
  {Physical Review B}\ }\textbf {\bibinfo {volume} {85}},\ \bibinfo {pages}
  {134428} (\bibinfo {year} {2012})}\BibitemShut {NoStop}%
\bibitem [{\citenamefont {Johnson}\ \emph {et~al.}(2012)\citenamefont
  {Johnson}, \citenamefont {de~Souza}, \citenamefont {Staub}, \citenamefont
  {Beaud}, \citenamefont {M{\"{o}}hr-Vorobeva}, \citenamefont {Ingold},
  \citenamefont {Caviezel}, \citenamefont {Scagnoli}, \citenamefont
  {Schlotter}, \citenamefont {Turner}, \citenamefont {Krupin}, \citenamefont
  {Lee}, \citenamefont {Chuang}, \citenamefont {Patthey}, \citenamefont
  {Moore}, \citenamefont {Lu}, \citenamefont {Yi}, \citenamefont {Kirchmann},
  \citenamefont {Trigo}, \citenamefont {Denes}, \citenamefont {Doering},
  \citenamefont {Hussain}, \citenamefont {Shen}, \citenamefont {Prabhakaran},\
  and\ \citenamefont {Boothroyd}}]{Johnson12}%
  \BibitemOpen
  \bibfield  {author} {\bibinfo {author} {\bibfnamefont {S.~L.}~\bibnamefont
  {Johnson}}, \bibinfo {author} {\bibfnamefont {R.~A.}~\bibnamefont {de~Souza}},
  \bibinfo {author} {\bibfnamefont {U.}~\bibnamefont {Staub}}, \bibinfo
  {author} {\bibfnamefont {P.}~\bibnamefont {Beaud}}, \bibinfo {author}
  {\bibfnamefont {E.}~\bibnamefont {M{\"{o}}hr-Vorobeva}}, \bibinfo {author}
  {\bibfnamefont {G.}~\bibnamefont {Ingold}}, \bibinfo {author} {\bibfnamefont
  {A.}~\bibnamefont {Caviezel}}, \bibinfo {author} {\bibfnamefont
  {V.}~\bibnamefont {Scagnoli}}, \bibinfo {author} {\bibfnamefont
  {W.~F.}~\bibnamefont {Schlotter}}, \bibinfo {author} {\bibfnamefont
  {J.~J.}~\bibnamefont {Turner}}, \bibinfo {author} {\bibfnamefont
  {O.}~\bibnamefont {Krupin}}, \bibinfo {author} {\bibfnamefont {W.-S.}\
  \bibnamefont {Lee}}, \bibinfo {author} {\bibfnamefont {Y.-D.}\ \bibnamefont
  {Chuang}}, \bibinfo {author} {\bibfnamefont {L.}~\bibnamefont {Patthey}},
  \bibinfo {author} {\bibfnamefont {R.~G.}~\bibnamefont {Moore}}, \bibinfo
  {author} {\bibfnamefont {D.}~\bibnamefont {Lu}}, \bibinfo {author}
  {\bibfnamefont {M.}~\bibnamefont {Yi}}, \bibinfo {author} {\bibfnamefont
  {P.~S.}~\bibnamefont {Kirchmann}}, \bibinfo {author} {\bibfnamefont
  {M.}~\bibnamefont {Trigo}}, \bibinfo {author} {\bibfnamefont
  {P.}~\bibnamefont {Denes}}, \bibinfo {author} {\bibfnamefont
  {D.}~\bibnamefont {Doering}}, \bibinfo {author} {\bibfnamefont
  {Z.}~\bibnamefont {Hussain}}, \bibinfo {author} {\bibfnamefont {Z.-X.}\
  \bibnamefont {Shen}}, \bibinfo {author} {\bibfnamefont {D.}~\bibnamefont
  {Prabhakaran}}, \ and\ \bibinfo {author} {\bibfnamefont {A.~T.}~\bibnamefont
  {Boothroyd}},\ }\href {\doibase 10.1103/PhysRevLett.108.037203} {\bibfield
  {journal} {\bibinfo  {journal} {Physical Review Letters}\ }\textbf {\bibinfo
  {volume} {108}},\ \bibinfo {pages} {037203} (\bibinfo {year}
  {2012})}\BibitemShut {NoStop}%
\bibitem [{\citenamefont {Villarreal}\ \emph {et~al.}(2012)\citenamefont
  {Villarreal}, \citenamefont {Quirion}, \citenamefont {Plumer}, \citenamefont
  {Poirier}, \citenamefont {Usui},\ and\ \citenamefont
  {Kimura}}]{Villarreal12}%
  \BibitemOpen
  \bibfield  {author} {\bibinfo {author} {\bibfnamefont {R.}~\bibnamefont
  {Villarreal}}, \bibinfo {author} {\bibfnamefont {G.}~\bibnamefont {Quirion}},
  \bibinfo {author} {\bibfnamefont {M.~L.}\ \bibnamefont {Plumer}}, \bibinfo
  {author} {\bibfnamefont {M.}~\bibnamefont {Poirier}}, \bibinfo {author}
  {\bibfnamefont {T.}~\bibnamefont {Usui}}, \ and\ \bibinfo {author}
  {\bibfnamefont {T.}~\bibnamefont {Kimura}},\ }\href {\doibase
  10.1103/PhysRevLett.109.167206} {\bibfield  {journal} {\bibinfo  {journal}
  {Physical Review Letters}\ }\textbf {\bibinfo {volume} {109}},\ \bibinfo
  {pages} {167206} (\bibinfo {year} {2012})}\BibitemShut {NoStop}%
\bibitem [{\citenamefont {Forsyth}\ \emph {et~al.}(1988)\citenamefont
  {Forsyth}, \citenamefont {Brown},\ and\ \citenamefont {Wanklyn}}]{Forsyth88}%
  \BibitemOpen
  \bibfield  {author} {\bibinfo {author} {\bibfnamefont {J.~B.}\ \bibnamefont
  {Forsyth}}, \bibinfo {author} {\bibfnamefont {P.~J.}\ \bibnamefont {Brown}},
  \ and\ \bibinfo {author} {\bibfnamefont {B.~M.}\ \bibnamefont {Wanklyn}},\
  }\href {\doibase 10.1088/0022-3719/21/15/023} {\bibfield  {journal} {\bibinfo
   {journal} {Journal of Physics C: Solid State Physics}\ }\textbf {\bibinfo
  {volume} {21}},\ \bibinfo {pages} {2917} (\bibinfo {year}
  {1988})}\BibitemShut {NoStop}%
\bibitem [{\citenamefont {Yang}\ \emph {et~al.}(1989)\citenamefont {Yang},
  \citenamefont {Thurston}, \citenamefont {Tranquada},\ and\ \citenamefont
  {Shirane}}]{Yang89}%
  \BibitemOpen
  \bibfield  {author} {\bibinfo {author} {\bibfnamefont {B.~X.}~\bibnamefont
  {Yang}}, \bibinfo {author} {\bibfnamefont {T.~R.}~\bibnamefont {Thurston}},
  \bibinfo {author} {\bibfnamefont {J.M.}~\bibnamefont {Tranquada}}, \ and\
  \bibinfo {author} {\bibfnamefont {G.}~\bibnamefont {Shirane}},\ }\href
  {\doibase 10.1103/PhysRevB.39.4343} {\bibfield  {journal} {\bibinfo
  {journal} {Physical Review B}\ }\textbf {\bibinfo {volume} {39}},\ \bibinfo
  {pages} {4343} (\bibinfo {year} {1989})}\BibitemShut {NoStop}%
\bibitem [{\citenamefont {Brown}\ \emph {et~al.}(1991)\citenamefont {Brown},
  \citenamefont {Chattopadhyay}, \citenamefont {Forsyth},\ and\ \citenamefont
  {Nunez}}]{Brown91}%
  \BibitemOpen
  \bibfield  {author} {\bibinfo {author} {\bibfnamefont {P.~J.}\ \bibnamefont
  {Brown}}, \bibinfo {author} {\bibfnamefont {T.}~\bibnamefont
  {Chattopadhyay}}, \bibinfo {author} {\bibfnamefont {J.~B.}\ \bibnamefont
  {Forsyth}}, \ and\ \bibinfo {author} {\bibfnamefont {V.}~\bibnamefont
  {Nunez}},\ }\href {\doibase 10.1088/0953-8984/3/23/016} {\bibfield  {journal}
  {\bibinfo  {journal} {Journal of Physics: Condensed Matter}\ }\textbf
  {\bibinfo {volume} {3}},\ \bibinfo {pages} {4281} (\bibinfo {year}
  {1991})}\BibitemShut {NoStop}%
\bibitem [{\citenamefont {Giovannetti}\ \emph
  {et~al.}(2011{\natexlab{a}})\citenamefont {Giovannetti}, \citenamefont
  {Kumar}, \citenamefont {Stroppa}, \citenamefont {van~den Brink},
  \citenamefont {Picozzi},\ and\ \citenamefont {Lorenzana}}]{Giovannetti11}%
  \BibitemOpen
  \bibfield  {author} {\bibinfo {author} {\bibfnamefont {G.}~\bibnamefont
  {Giovannetti}}, \bibinfo {author} {\bibfnamefont {S.}~\bibnamefont {Kumar}},
  \bibinfo {author} {\bibfnamefont {A.}~\bibnamefont {Stroppa}}, \bibinfo
  {author} {\bibfnamefont {J.}~\bibnamefont {van~den Brink}}, \bibinfo {author}
  {\bibfnamefont {S.}~\bibnamefont {Picozzi}}, \ and\ \bibinfo {author}
  {\bibfnamefont {J.}~\bibnamefont {Lorenzana}},\ }\href {\doibase
  10.1103/PhysRevLett.106.026401} {\bibfield  {journal} {\bibinfo  {journal}
  {Physical Review Letters}\ }\textbf {\bibinfo {volume} {106}},\ \bibinfo
  {pages} {026401} (\bibinfo {year} {2011}{\natexlab{a}})}\BibitemShut
  {NoStop}%
\bibitem [{\citenamefont {Jin}\ \emph {et~al.}(2012)\citenamefont {Jin},
  \citenamefont {Cao}, \citenamefont {Guo},\ and\ \citenamefont {He}}]{Jin12}%
  \BibitemOpen
  \bibfield  {author} {\bibinfo {author} {\bibfnamefont {G.}~\bibnamefont
  {Jin}}, \bibinfo {author} {\bibfnamefont {K.}~\bibnamefont {Cao}}, \bibinfo
  {author} {\bibfnamefont {G.-C.}\ \bibnamefont {Guo}}, \ and\ \bibinfo
  {author} {\bibfnamefont {L.}~\bibnamefont {He}},\ }\href {\doibase
  10.1103/PhysRevLett.108.187205} {\bibfield  {journal} {\bibinfo  {journal}
  {Physical Review Letters}\ }\textbf {\bibinfo {volume} {108}},\ \bibinfo
  {pages} {187205} (\bibinfo {year} {2012})}\BibitemShut {NoStop}%
\bibitem [{\citenamefont {Kaplan}(2009)}]{Kaplan09}%
  \BibitemOpen
  \bibfield  {author} {\bibinfo {author} {\bibfnamefont {T.~A.}~\bibnamefont
  {Kaplan}},\ }\href {\doibase 10.1103/PhysRevB.80.012407} {\bibfield
  {journal} {\bibinfo  {journal} {Physical Review B}\ }\textbf {\bibinfo
  {volume} {80}},\ \bibinfo {pages} {012407} (\bibinfo {year}
  {2009})}\BibitemShut {NoStop}%
\bibitem [{\citenamefont {Henley}(1989)}]{Henley89}%
  \BibitemOpen
  \bibfield  {author} {\bibinfo {author} {\bibfnamefont {C.~L.}~\bibnamefont
  {Henley}},\ }\href {\doibase 10.1103/PhysRevLett.62.2056} {\bibfield
  {journal} {\bibinfo  {journal} {Physical Review Letters}\ }\textbf {\bibinfo
  {volume} {62}},\ \bibinfo {pages} {2056} (\bibinfo {year}
  {1989})}\BibitemShut {NoStop}%
\bibitem [{\citenamefont {Yablonskii}(1990)}]{Yablonskii90}%
  \BibitemOpen
  \bibfield  {author} {\bibinfo {author} {\bibfnamefont {D.}~\bibnamefont
  {Yablonskii}},\ }\href {\doibase 10.1016/0921-4534(90)90257-F} {\bibfield
  {journal} {\bibinfo  {journal} {Physica C: Superconductivity}\ }\textbf
  {\bibinfo {volume} {171}},\ \bibinfo {pages} {454} (\bibinfo {year}
  {1990})}\BibitemShut {NoStop}%
\bibitem [{\citenamefont {Filippetti}\ and\ \citenamefont
  {Fiorentini}(2005)}]{Filippetti05}%
  \BibitemOpen
  \bibfield  {author} {\bibinfo {author} {\bibfnamefont {A.}~\bibnamefont
  {Filippetti}}\ and\ \bibinfo {author} {\bibfnamefont {V.}~\bibnamefont
  {Fiorentini}},\ }\href {\doibase 10.1103/PhysRevLett.95.086405} {\bibfield
  {journal} {\bibinfo  {journal} {Physical Review Letters}\ }\textbf {\bibinfo
  {volume} {95}},\ \bibinfo {pages} {086405} (\bibinfo {year}
  {2005})}\BibitemShut {NoStop}%
\bibitem [{\citenamefont {Rocquefelte}\ \emph {et~al.}(2010)\citenamefont
  {Rocquefelte}, \citenamefont {Whangbo}, \citenamefont {Villesuzanne},
  \citenamefont {Jobic}, \citenamefont {Tran}, \citenamefont {Schwarz},\ and\
  \citenamefont {Blaha}}]{Rocquefelte10}%
  \BibitemOpen
  \bibfield  {author} {\bibinfo {author} {\bibfnamefont {X.}~\bibnamefont
  {Rocquefelte}}, \bibinfo {author} {\bibfnamefont {M.-H.}\ \bibnamefont
  {Whangbo}}, \bibinfo {author} {\bibfnamefont {A.}~\bibnamefont
  {Villesuzanne}}, \bibinfo {author} {\bibfnamefont {S.}~\bibnamefont {Jobic}},
  \bibinfo {author} {\bibfnamefont {F.}~\bibnamefont {Tran}}, \bibinfo {author}
  {\bibfnamefont {K.}~\bibnamefont {Schwarz}}, \ and\ \bibinfo {author}
  {\bibfnamefont {P.}~\bibnamefont {Blaha}},\ }\href {\doibase
  10.1088/0953-8984/22/4/045502} {\bibfield  {journal} {\bibinfo  {journal}
  {Journal of Physics: Condensed Matter}\ }\textbf {\bibinfo {volume} {22}},\
  \bibinfo {pages} {045502} (\bibinfo {year} {2010})}\BibitemShut {NoStop}%
\bibitem [{\citenamefont {Rocquefelte}\ \emph {et~al.}(2012)\citenamefont
  {Rocquefelte}, \citenamefont {Schwarz},\ and\ \citenamefont
  {Blaha}}]{Rocquefelte12}%
  \BibitemOpen
  \bibfield  {author} {\bibinfo {author} {\bibfnamefont {X.}~\bibnamefont
  {Rocquefelte}}, \bibinfo {author} {\bibfnamefont {K.}~\bibnamefont
  {Schwarz}}, \ and\ \bibinfo {author} {\bibfnamefont {P.}~\bibnamefont
  {Blaha}},\ }\href {\doibase 10.1038/srep00759} {\bibfield  {journal}
  {\bibinfo  {journal} {Scientific Reports}\ }\textbf {\bibinfo {volume} {2}},\
  \bibinfo {pages} {759} (\bibinfo {year} {2012})}\BibitemShut {NoStop}%
\bibitem [{\citenamefont {Kresse}\ and\ \citenamefont
  {Furthm{\"{u}}ller}(1996)}]{Kresse96}%
  \BibitemOpen
  \bibfield  {author} {\bibinfo {author} {\bibfnamefont {G.}~\bibnamefont
  {Kresse}}\ and\ \bibinfo {author} {\bibfnamefont {J.}~\bibnamefont
  {Furthm{\"{u}}ller}},\ }\href {\doibase 10.1016/0927-0256(96)00008-0}
  {\bibfield  {journal} {\bibinfo  {journal} {Computational Materials Science}\
  }\textbf {\bibinfo {volume} {6}},\ \bibinfo {pages} {15} (\bibinfo {year}
  {1996})}\BibitemShut {NoStop}%
\bibitem [{\citenamefont {Anisimov}\ \emph {et~al.}(1999)\citenamefont
  {Anisimov}, \citenamefont {Aryasetiawan},\ and\ \citenamefont
  {Lichtenstein}}]{Anisimov97}%
  \BibitemOpen
  \bibfield  {author} {\bibinfo {author} {\bibfnamefont {V.~I.}\ \bibnamefont
  {Anisimov}}, \bibinfo {author} {\bibfnamefont {F.}~\bibnamefont
  {Aryasetiawan}}, \ and\ \bibinfo {author} {\bibfnamefont {A.~I.}\
  \bibnamefont {Lichtenstein}},\ }\href {\doibase 10.1088/0953-8984/9/4/002}
  {\bibfield  {journal} {\bibinfo  {journal} {Journal of Physics: Condensed
  Matter}\ }\textbf {\bibinfo {volume} {9}},\ \bibinfo {pages} {767} (\bibinfo
  {year} {1999})}\BibitemShut {NoStop}%
\bibitem [{\citenamefont {Dudarev}\ \emph {et~al.}(1998)\citenamefont
  {Dudarev}, \citenamefont {Botton}, \citenamefont {Savrasov}, \citenamefont
  {Humphreys},\ and\ \citenamefont {Sutton}}]{Dudarev98}%
  \BibitemOpen
  \bibfield  {author} {\bibinfo {author} {\bibfnamefont {S.~L.}\ \bibnamefont
  {Dudarev}}, \bibinfo {author} {\bibfnamefont {G.~A.}\ \bibnamefont {Botton}},
  \bibinfo {author} {\bibfnamefont {S.~Y.}\ \bibnamefont {Savrasov}}, \bibinfo
  {author} {\bibfnamefont {C.~J.}\ \bibnamefont {Humphreys}}, \ and\ \bibinfo
  {author} {\bibfnamefont {A.~P.}\ \bibnamefont {Sutton}},\ }\href@noop {}
  {\bibfield  {journal} {\bibinfo  {journal} {Physical Review B}\ }\textbf
  {\bibinfo {volume} {57}},\ \bibinfo {pages} {1505} (\bibinfo {year}
  {1998})}\BibitemShut {NoStop}%
\bibitem [{\citenamefont {Perdew}\ \emph {et~al.}(1992)\citenamefont {Perdew},
  \citenamefont {Jackson}, \citenamefont {Pederson}, \citenamefont {Singh},\
  and\ \citenamefont {Fiolhais}}]{Perdew92}%
  \BibitemOpen
  \bibfield  {author} 
{\bibinfo {author} {\bibfnamefont {J.~P.}\ \bibnamefont{Perdew}}, 
 \bibinfo {author} {\bibfnamefont {J.~A.}\ \bibnamefont {Chevary}},
 \bibinfo {author} {\bibfnamefont {S.~H.}\ \bibnamefont {Vosko}},
 \bibinfo {author} {\bibfnamefont {K.~A.}\ \bibnamefont {Jackson}},
 \bibinfo {author} {\bibfnamefont {M.~R.}\ \bibnamefont {Pederson}}, \bibinfo
  {author} {\bibfnamefont {D.~J.}\ \bibnamefont {Singh}}, \ and\ \bibinfo
  {author} {\bibfnamefont {C.}~\bibnamefont {Fiolhais}},\ }\href {\doibase
  10.1103/PhysRevB.46.6671} {\bibfield  {journal} {\bibinfo  {journal}
  {Physical Review B}\ }\textbf {\bibinfo {volume} {46}},\ \bibinfo {pages}
  {6671} (\bibinfo {year} {1992})}\BibitemShut {NoStop}%
\bibitem [{\citenamefont {Perdew}\ \emph {et~al.}(1996)\citenamefont {Perdew},
  \citenamefont {Burke},\ and\ \citenamefont {Ernzerhof}}]{Perdew96}%
  \BibitemOpen
  \bibfield  {author} {\bibinfo {author} {\bibfnamefont {J.~P.}\ \bibnamefont
  {Perdew}}, \bibinfo {author} {\bibfnamefont {K.}~\bibnamefont {Burke}}, \
  and\ \bibinfo {author} {\bibfnamefont {M.}~\bibnamefont {Ernzerhof}},\ }\href
  {\doibase 10.1103/PhysRevLett.77.3865} {\bibfield  {journal} {\bibinfo
  {journal} {Physical Review Letters}\ }\textbf {\bibinfo {volume} {77}},\
  \bibinfo {pages} {3865} (\bibinfo {year} {1996})}\BibitemShut {NoStop}%
\bibitem [{\citenamefont {{\u{A}}sbrink}\ and\ \citenamefont
  {Norrby}(1970)}]{Asbrink70}%
  \BibitemOpen
  \bibfield  {author} {\bibinfo {author} {\bibfnamefont {S.}~\bibnamefont
  {{\u{A}}sbrink}}\ and\ \bibinfo {author} {\bibfnamefont {L.~J.}\ \bibnamefont
  {Norrby}},\ }\href {\doibase 10.1107/S0567740870001838} {\bibfield  {journal}
  {\bibinfo  {journal} {Acta Crystallographica Section B Structural
  Crystallography and Crystal Chemistry}\ }\textbf {\bibinfo {volume} {26}},\
  \bibinfo {pages} {8} (\bibinfo {year} {1970})}\BibitemShut {NoStop}%
\bibitem [{\citenamefont {Rocquefelte}\ \emph {et~al.}(2011)\citenamefont
  {Rocquefelte}, \citenamefont {Schwarz},\ and\ \citenamefont
  {Blaha}}]{Rocquefelte11}%
  \BibitemOpen
  \bibfield  {author} {\bibinfo {author} {\bibfnamefont {X.}~\bibnamefont
  {Rocquefelte}}, \bibinfo {author} {\bibfnamefont {K.}~\bibnamefont
  {Schwarz}}, \ and\ \bibinfo {author} {\bibfnamefont {P.}~\bibnamefont
  {Blaha}},\ }\href {\doibase 10.1103/PhysRevLett.107.239701} {\bibfield
  {journal} {\bibinfo  {journal} {Physical Review Letters}\ }\textbf {\bibinfo
  {volume} {107}},\ \bibinfo {pages} {239701} (\bibinfo {year}
  {2011})}\BibitemShut {NoStop}%
\bibitem [{\citenamefont {Giovannetti}\ \emph
  {et~al.}(2011{\natexlab{b}})\citenamefont {Giovannetti}, \citenamefont
  {Kumar}, \citenamefont {Stroppa}, \citenamefont {Balestieri}, \citenamefont
  {van~den Brink}, \citenamefont {Picozzi},\ and\ \citenamefont
  {Lorenzana}}]{Giovannetti11-1}%
  \BibitemOpen
  \bibfield  {author} {\bibinfo {author} {\bibfnamefont {G.}~\bibnamefont
  {Giovannetti}}, \bibinfo {author} {\bibfnamefont {S.}~\bibnamefont {Kumar}},
  \bibinfo {author} {\bibfnamefont {A.}~\bibnamefont {Stroppa}}, \bibinfo
  {author} {\bibfnamefont {M.}~\bibnamefont {Balestieri}}, \bibinfo {author}
  {\bibfnamefont {J.}~\bibnamefont {van~den Brink}}, \bibinfo {author}
  {\bibfnamefont {S.}~\bibnamefont {Picozzi}}, \ and\ \bibinfo {author}
  {\bibfnamefont {J.}~\bibnamefont {Lorenzana}},\ }\href {\doibase
  10.1103/PhysRevLett.107.239702} {\bibfield  {journal} {\bibinfo  {journal}
  {Physical Review Letters}\ }\textbf {\bibinfo {volume} {107}},\ \bibinfo
  {pages} {239702} (\bibinfo {year} {2011}{\natexlab{b}})}\BibitemShut
  {NoStop}%
\bibitem [{\citenamefont {Goodenough}(1955)}]{Goodenough55}%
  \BibitemOpen
  \bibfield  {author} {\bibinfo {author} {\bibfnamefont {J.~B.}\ \bibnamefont
  {Goodenough}},\ }\href {\doibase 10.1103/PhysRev.100.564} {\bibfield
  {journal} {\bibinfo  {journal} {Physical Review}\ }\textbf {\bibinfo {volume}
  {100}},\ \bibinfo {pages} {564} (\bibinfo {year} {1955})}\BibitemShut
  {NoStop}%
\bibitem [{\citenamefont {Kanamori}(1963)}]{Kanamori63}%
  \BibitemOpen
  \bibfield  {author} {\bibinfo {author} {\bibfnamefont {J.}~\bibnamefont
  {Kanamori}},\ }\href {\doibase 10.1143/PTP.30.275} {\bibfield  {journal}
  {\bibinfo  {journal} {Progress of Theoretical Physics}\ }\textbf {\bibinfo
  {volume} {30}},\ \bibinfo {pages} {275} (\bibinfo {year} {1963})}\BibitemShut
  {NoStop}%
\bibitem [{\citenamefont {Nikuni}\ and\ \citenamefont
  {Jacobs}(1998)}]{Nikuni98}%
  \BibitemOpen
  \bibfield  {author} {\bibinfo {author} {\bibfnamefont {T.}~\bibnamefont
  {Nikuni}}\ and\ \bibinfo {author} {\bibfnamefont {A.~E.}~\bibnamefont
  {Jacobs}},\ }\href {\doibase 10.1103/PhysRevB.57.5205} {\bibfield  {journal}
  {\bibinfo  {journal} {Physical Review B}\ }\textbf {\bibinfo {volume} {57}},\
  \bibinfo {pages} {5205} (\bibinfo {year} {1998})}\BibitemShut {NoStop}%
\bibitem [{\citenamefont {Hukushima}\ and\ \citenamefont
  {Nemoto}(1996)}]{Hukushima96}%
  \BibitemOpen
  \bibfield  {author} {\bibinfo {author} {\bibfnamefont {K.}~\bibnamefont
  {Hukushima}}\ and\ \bibinfo {author} {\bibfnamefont {K.}~\bibnamefont
  {Nemoto}},\ }\href {\doibase 10.1143/JPSJ.65.1604} {\bibfield  {journal}
  {\bibinfo  {journal} {Journal of the Physics Society Japan}\ }\textbf
  {\bibinfo {volume} {65}},\ \bibinfo {pages} {1604} (\bibinfo {year}
  {1996})}\BibitemShut {NoStop}%
\bibitem [{\citenamefont {Martin-Mayor}(2007)}]{Martin-Mayor07}%
  \BibitemOpen
  \bibfield  {author} {\bibinfo {author} {\bibfnamefont {V.}~\bibnamefont
  {Martin-Mayor}},\ }\href {\doibase 10.1103/PhysRevLett.98.137207} {\bibfield
  {journal} {\bibinfo  {journal} {Physical Review Letters}\ }\textbf {\bibinfo
  {volume} {98}},\ \bibinfo {pages} {137207} (\bibinfo {year}
  {2007})}\BibitemShut {NoStop}%
\bibitem [{\citenamefont {Katsura}\ \emph {et~al.}(2005)\citenamefont
  {Katsura}, \citenamefont {Nagaosa},\ and\ \citenamefont
  {Balatsky}}]{Katsura05}%
  \BibitemOpen
  \bibfield  {author} {\bibinfo {author} {\bibfnamefont {H.}~\bibnamefont
  {Katsura}}, \bibinfo {author} {\bibfnamefont {N.}~\bibnamefont {Nagaosa}}, \
  and\ \bibinfo {author} {\bibfnamefont {A.~V.}\ \bibnamefont {Balatsky}},\
  }\href {\doibase 10.1103/PhysRevLett.95.057205} {\bibfield  {journal}
  {\bibinfo  {journal} {Physical Review Letters}\ }\textbf {\bibinfo {volume}
  {95}},\ \bibinfo {pages} {057205} (\bibinfo {year} {2005})}\BibitemShut
  {NoStop}%
\bibitem [{\citenamefont {Chatterji}\ \emph {et~al.}(2005)\citenamefont
  {Chatterji}, \citenamefont {Brown},\ and\ \citenamefont
  {Forsyth}}]{Chatterji05}%
  \BibitemOpen
  \bibfield  {author} {\bibinfo {author} {\bibfnamefont {T.}~\bibnamefont
  {Chatterji}}, \bibinfo {author} {\bibfnamefont {P.~J.}\ \bibnamefont
  {Brown}}, \ and\ \bibinfo {author} {\bibfnamefont {J.~B.}\ \bibnamefont
  {Forsyth}},\ }\href {\doibase 10.1088/0953-8984/17/40/008} {\bibfield
  {journal} {\bibinfo  {journal} {Journal of Physics: Condensed Matter}\
  }\textbf {\bibinfo {volume} {17}},\ \bibinfo {pages} {S3057} (\bibinfo {year}
  {2005})}\BibitemShut {NoStop}%
\bibitem [{\citenamefont {Wang}\ \emph {et~al.}(2006)\citenamefont {Wang},
  \citenamefont {Maxisch},\ and\ \citenamefont {Ceder}}]{Wang06-1}%
  \BibitemOpen
  \bibfield  {author} {\bibinfo {author} {\bibfnamefont {L.}~\bibnamefont
  {Wang}}, \bibinfo {author} {\bibfnamefont {T.}~\bibnamefont {Maxisch}}, \
  and\ \bibinfo {author} {\bibfnamefont {G.}~\bibnamefont {Ceder}},\ }\href
  {\doibase 10.1103/PhysRevB.73.195107} {\bibfield  {journal} {\bibinfo
  {journal} {Physical Review B}\ }\textbf {\bibinfo {volume} {73}},\ \bibinfo
  {pages} {195107} (\bibinfo {year} {2006})}\BibitemShut {NoStop}%
\bibitem [{\citenamefont {Rohrbach}\ \emph {et~al.}(2004)\citenamefont
  {Rohrbach}, \citenamefont {Hafner},\ and\ \citenamefont
  {Kresse}}]{Rohrbach04}%
  \BibitemOpen
  \bibfield  {author} {\bibinfo {author} {\bibfnamefont {A.}~\bibnamefont
  {Rohrbach}}, \bibinfo {author} {\bibfnamefont {J.}~\bibnamefont {Hafner}}, \
  and\ \bibinfo {author} {\bibfnamefont {G.}~\bibnamefont {Kresse}},\ }\href
  {\doibase 10.1103/PhysRevB.69.075413} {\bibfield  {journal} {\bibinfo
  {journal} {Physical Review B}\ }\textbf {\bibinfo {volume} {69}},\ \bibinfo
  {pages} {075413} (\bibinfo {year} {2004})}\BibitemShut {NoStop}%
\bibitem [{\citenamefont {Chen}\ \emph {et~al.}(2011)\citenamefont {Chen},
  \citenamefont {Wu},\ and\ \citenamefont {Selloni}}]{Chen11-1}%
  \BibitemOpen
  \bibfield  {author} {\bibinfo {author} {\bibfnamefont {J.}~\bibnamefont
  {Chen}}, \bibinfo {author} {\bibfnamefont {X.}~\bibnamefont {Wu}}, \ and\
  \bibinfo {author} {\bibfnamefont {A.}~\bibnamefont {Selloni}},\ }\href
  {\doibase 10.1103/PhysRevB.83.245204} {\bibfield  {journal} {\bibinfo
  {journal} {Physical Review B}\ }\textbf {\bibinfo {volume} {83}},\ \bibinfo
  {pages} {245204} (\bibinfo {year} {2011})}\BibitemShut {NoStop}%
\bibitem{note}In order to
estimate the total magnetization of ions the magnetization was
integrated within a sphere. For Cu the radius was 1.65~\AA, for Cd and
Zn the radius was set to 1.300~\AA, for Mg 1.524~\AA, for Co
1.45~\AA~, and for Ni 1.43~\AA. 

\bibitem [{\citenamefont {Braden}\ \emph {et~al.}(1996)\citenamefont {Braden},
  \citenamefont {Wilkendorf}, \citenamefont {Lorenzana}, \citenamefont
  {A{\"{i}}n}, \citenamefont {McIntyre}, \citenamefont {Behruzi}, \citenamefont
  {Heger}, \citenamefont {Dhalenne},\ and\ \citenamefont
  {Revcolevschi}}]{Braden96}%
  \BibitemOpen
  \bibfield  {author} {\bibinfo {author} {\bibfnamefont {M.}~\bibnamefont
  {Braden}}, \bibinfo {author} {\bibfnamefont {G.}~\bibnamefont {Wilkendorf}},
  \bibinfo {author} {\bibfnamefont {J.}~\bibnamefont {Lorenzana}}, \bibinfo
  {author} {\bibfnamefont {M.}~\bibnamefont {A{\"{i}}n}}, \bibinfo {author}
  {\bibfnamefont {G.~J.}~\bibnamefont {McIntyre}}, \bibinfo {author}
  {\bibfnamefont {M.}~\bibnamefont {Behruzi}}, \bibinfo {author} {\bibfnamefont
  {G.}~\bibnamefont {Heger}}, \bibinfo {author} {\bibfnamefont
  {G.}~\bibnamefont {Dhalenne}}, \ and\ \bibinfo {author} {\bibfnamefont
  {A.}~\bibnamefont {Revcolevschi}},\ }\href {\doibase
  10.1103/PhysRevB.54.1105} {\bibfield  {journal} {\bibinfo  {journal}
  {Physical Review B}\ }\textbf {\bibinfo {volume} {54}},\ \bibinfo {pages}
  {1105} (\bibinfo {year} {1996})}\BibitemShut {NoStop}%
\bibitem [{\citenamefont {Prabhakaran}\ and\ \citenamefont
  {Boothroyd}(2003)}]{Prabhakaran03}%
  \BibitemOpen
  \bibfield  {author} {\bibinfo {author} {\bibfnamefont {D.}~\bibnamefont
  {Prabhakaran}}\ and\ \bibinfo {author} {\bibfnamefont {A.}~\bibnamefont
  {Boothroyd}},\ }\href {\doibase 10.1016/S0022-0248(02)02230-3} {\bibfield
  {journal} {\bibinfo  {journal} {Journal of Crystal Growth}\ }\textbf
  {\bibinfo {volume} {250}},\ \bibinfo {pages} {77} (\bibinfo {year}
  {2003})}\BibitemShut {NoStop}%
\bibitem [{Note1()}]{Note1}%
  \BibitemOpen
  \bibinfo {note} {Previous results for the magnetic anomalies at 5\% Zn doping
  are in well agreement for $T_{\protect \textrm {N2}}$ but yield a smaller
  widening of the window ($T_{\protect \textrm {N1(5\%)}}/T_{\protect \textrm
  {N2,(0)}}=0.74$) \cite {Prabhakaran03}. Reexamination of these samples showed
  that they where weakly conducting as opposed to the samples used in Fig.~\ref
  {fig:phasediag} which where insulating.}\BibitemShut {Stop}%
\bibitem [{\citenamefont {Rocquefelte}\ \emph {et~al.}(2013)\citenamefont
  {Rocquefelte}, \citenamefont {Schwarz}, \citenamefont {Blaha}, \citenamefont
  {Kumar},\ and\ \citenamefont {van~den Brink}}]{Rocquefelte13}%
  \BibitemOpen
  \bibfield  {author} {\bibinfo {author} {\bibfnamefont {X.}~\bibnamefont
  {Rocquefelte}}, \bibinfo {author} {\bibfnamefont {K.}~\bibnamefont
  {Schwarz}}, \bibinfo {author} {\bibfnamefont {P.}~\bibnamefont {Blaha}},
  \bibinfo {author} {\bibfnamefont {S.}~\bibnamefont {Kumar}}, \ and\ \bibinfo
  {author} {\bibfnamefont {J.}~\bibnamefont {van~den Brink}},\ }\href {\doibase
  10.1038/ncomms3511} {\bibfield  {journal} {\bibinfo  {journal} {Nature
  Communications}\ }\textbf {\bibinfo {volume} {4}},\ \bibinfo {pages} {2511}
  (\bibinfo {year} {2013})}\BibitemShut {NoStop}%
\bibitem [{Note2()}]{Note2}%
  \BibitemOpen
  \bibinfo {note} {We have checked with powder x-ray diffraction that the
  volume does not decrease with Zn doping.}\BibitemShut {Stop}%
\bibitem [{\citenamefont {Cheong}\ and\ \citenamefont
  {Mostovoy}(2007)}]{Cheong07}%
  \BibitemOpen
  \bibfield  {author} {\bibinfo {author} {\bibfnamefont {S.-W.}\ \bibnamefont
  {Cheong}}\ and\ \bibinfo {author} {\bibfnamefont {M.}~\bibnamefont
  {Mostovoy}},\ }\href {\doibase 10.1038/nmat1804} {\bibfield  {journal}
  {\bibinfo  {journal} {Nature Materials}\ }\textbf {\bibinfo {volume} {6}},\
  \bibinfo {pages} {13} (\bibinfo {year} {2007})}\BibitemShut {NoStop}%
\bibitem [{\citenamefont {Gukasov}\ \emph {et~al.}(1998)\citenamefont
  {Gukasov}, \citenamefont {Br{\"{u}}ckel}, \citenamefont {Dorner},
  \citenamefont {Plakhty}, \citenamefont {Prandl}, \citenamefont {Shender},\
  and\ \citenamefont {Smirnov}}]{Gukasov88}%
  \BibitemOpen
  \bibfield  {author} {\bibinfo {author} {\bibfnamefont {A.~G.}\ \bibnamefont
  {Gukasov}}, \bibinfo {author} {\bibfnamefont {T.}~\bibnamefont
  {Br{\"{u}}ckel}}, \bibinfo {author} {\bibfnamefont {B.}~\bibnamefont
  {Dorner}}, \bibinfo {author} {\bibfnamefont {V.~P.}\ \bibnamefont {Plakhty}},
  \bibinfo {author} {\bibfnamefont {W.}~\bibnamefont {Prandl}}, \bibinfo
  {author} {\bibfnamefont {E.~F.}\ \bibnamefont {Shender}}, \ and\ \bibinfo
  {author} {\bibfnamefont {O.~P.}\ \bibnamefont {Smirnov}},\ }\href {\doibase
  10.1209/0295-5075/7/1/014} {\bibfield  {journal} {\bibinfo  {journal}
  {Europhysics Letters (EPL)}\ }\textbf {\bibinfo {volume} {7}},\ \bibinfo
  {pages} {83} (\bibinfo {year} {1998})}\BibitemShut {NoStop}%
\bibitem [{\citenamefont {Giebultowicz}\ \emph {et~al.}(1993)\citenamefont
  {Giebultowicz}, \citenamefont {Kl{\l}osowski}, \citenamefont {Samarth},
  \citenamefont {Luo}, \citenamefont {Furdyna},\ and\ \citenamefont
  {Rhyne}}]{Giebultowicz93}%
  \BibitemOpen
  \bibfield  {author} {\bibinfo {author} {\bibfnamefont {T.~M.}~\bibnamefont
  {Giebultowicz}}, \bibinfo {author} {\bibfnamefont {P.}~\bibnamefont
  {K{\l}osowski}}, \bibinfo {author} {\bibfnamefont {N.}~\bibnamefont {Samarth}},
  \bibinfo {author} {\bibfnamefont {H.}~\bibnamefont {Luo}}, \bibinfo {author}
  {\bibfnamefont {J.~K.}~\bibnamefont {Furdyna}}, \ and\ \bibinfo {author}
  {\bibfnamefont {J.~J.}~\bibnamefont {Rhyne}},\ }\href {\doibase
  10.1103/PhysRevB.48.12817} {\bibfield  {journal} {\bibinfo  {journal}
  {Physical Review B}\ }\textbf {\bibinfo {volume} {48}},\ \bibinfo {pages}
  {12817} (\bibinfo {year} {1993})}\BibitemShut {NoStop}%
\bibitem [{\citenamefont {Capriotti}\ \emph {et~al.}(2004)\citenamefont
  {Capriotti}, \citenamefont {Fubini}, \citenamefont {Roscilde},\ and\
  \citenamefont {Tognetti}}]{Capriotti04}%
  \BibitemOpen
  \bibfield  {author} {\bibinfo {author} {\bibfnamefont {L.}~\bibnamefont
  {Capriotti}}, \bibinfo {author} {\bibfnamefont {A.}~\bibnamefont {Fubini}},
  \bibinfo {author} {\bibfnamefont {T.}~\bibnamefont {Roscilde}}, \ and\
  \bibinfo {author} {\bibfnamefont {V.}~\bibnamefont {Tognetti}},\ }\href
  {\doibase 10.1103/PhysRevLett.92.157202} {\bibfield  {journal} {\bibinfo
  {journal} {Physical Review Letters}\ }\textbf {\bibinfo {volume} {92}},\
  \bibinfo {pages} {157202} (\bibinfo {year} {2004})}\BibitemShut {NoStop}%
\bibitem [{\citenamefont {Diep}(2005)}]{2005FrustratedDiep}%
  \BibitemOpen
  \bibinfo {editor} {\bibfnamefont {H.~T.}\ \bibnamefont {Diep}},\ ed.,\ \href
  {\doibase 10.1142/9789812567819} {\emph {\bibinfo {title} {Frustrated spin
  systems}}}\ (\bibinfo  {publisher} {World Scientific},\ \bibinfo {address}
  {Hackensack, NJ},\ \bibinfo {year} {2005})\BibitemShut {NoStop}%
\bibitem [{\citenamefont {Lacroix}\ \emph {et~al.}(2010)\citenamefont
  {Lacroix}, \citenamefont {Mila},\ and\ \citenamefont
  {Mendels}}]{2011FrustratedLacroix}%
  \BibitemOpen
  \bibinfo {editor} {\bibfnamefont {C.}~\bibnamefont {Lacroix}}, \bibinfo
  {editor} {\bibfnamefont {F.}~\bibnamefont {Mila}}, \ and\ \bibinfo {editor}
  {\bibfnamefont {P.}~\bibnamefont {Mendels}},\ eds.,\ \href {\doibase
  10.1007/978-3-642-10589-0} {\emph {\bibinfo {title} {Introduction to
  frustrated magnetism : materials, experiments, theory}}}\ (\bibinfo
  {publisher} {Springer},\ \bibinfo {address} {Berlin},\ \bibinfo {year}
  {2010})\BibitemShut {NoStop}%
\bibitem [{\citenamefont {Savary}\ \emph {et~al.}(2012)\citenamefont {Savary},
  \citenamefont {Ross}, \citenamefont {Gaulin}, \citenamefont {Ruff},\ and\
  \citenamefont {Balents}}]{Savary12}%
  \BibitemOpen
  \bibfield  {author} {\bibinfo {author} {\bibfnamefont {L.}~\bibnamefont
  {Savary}}, \bibinfo {author} {\bibfnamefont {K.~A.}\ \bibnamefont {Ross}},
  \bibinfo {author} {\bibfnamefont {B.~D.}\ \bibnamefont {Gaulin}}, \bibinfo
  {author} {\bibfnamefont {J.~P.~C.}\ \bibnamefont {Ruff}}, \ and\ \bibinfo
  {author} {\bibfnamefont {L.}~\bibnamefont {Balents}},\ }\href {\doibase
  10.1103/PhysRevLett.109.167201} {\bibfield  {journal} {\bibinfo  {journal}
  {Physical Review Letters}\ }\textbf {\bibinfo {volume} {109}},\ \bibinfo
  {pages} {167201} (\bibinfo {year} {2012})}\BibitemShut {NoStop}%
\end{thebibliography}

%

\end{document}